%% file: Time_Reversal.tex
\documentclass[journal,letterpaper,twoside]{IEEEtran}
\ifCLASSOPTIONcompsoc
   \usepackage[nocompress]{cite}
\else
   \usepackage[nospace]{cite}
\fi
\ifCLASSINFOpdf
   \usepackage[pdftex]{graphicx}
\else
   \usepackage[dvips]{graphicx}
\fi
\usepackage[cmex10]{amsmath}
\usepackage{array}
\ifCLASSOPTIONcompsoc
\usepackage[caption=false,font=normalsize,labelfont=sf,textfont=sf]{subfig}
\else
\usepackage[caption=false,font=footnotesize]{subfig}
\fi
\usepackage{dblfloatfix}
\usepackage{url}
\usepackage{wasysym}
\usepackage{blindtext}
\usepackage{amsthm,amsmath}
\usepackage{xfrac}
\usepackage{amssymb}

\newcommand{\ud}{\: \mathrm{d}}
\theoremstyle{remark} 
\theoremstyle{remark}

\newcommand{\vb}[1]{\mathbf{#1}}

\begin{document}
\title{Optimal Time-Reversed Wideband Signals for Distributed Sensing}

\author{Jerry~Kim$^{**\ddag }$, Margaret~Cheney$^{\dagger}$, Eric~Mokole$^*$ % <-this % stops a space 
\protect\\

\IEEEcompsocitemizethanks{\IEEEcompsocthanksitem \protect \\ $\ddag$ Ph.D. Student, Rensselaer Polytechnic Institute \protect\\
$*$ Consultant \protect\\ $**$ Tactical Electronic Warfare Division, Naval Research Laboratory, Washington DC 20375, USA \protect\\ 	$\dagger$ Department of Mathematics, Colorado State University \protect\\
\IEEEcompsocthanksitem E-mail: } %
\thanks{Manuscript written on \today}}


\markboth{IEEE TRANSACTIONS ON ANTENNA AND PROPAGATION,~Vol.~XX, No.~X, Month~xxxx}%
{Kim \MakeLowercase{\textit{et al.}}:Iterative Time-Reversed Wideband Signals}

%\IEEEpubid{\makebox[\columnwidth]{\hfill 0000--0000/00/\$00.00~\copyright~2012 IEEE}%


\input{./sections/abstractnwords}
\maketitle

\IEEEdisplaynontitleabstractindextext

% \fi
\IEEEpeerreviewmaketitle
%\IEEEpubidadjcol

\input{./sections/body}
\input{./sections/appendix}
\input{./sections/acknowledgment}
\input{./sections/reference}
\input{./sections/biography}
\end{document}

%% file: sections/abstractnwords.tex
\IEEEtitleabstractindextext{%
\begin{abstract}
This paper considers a distributed wave-based sensing system that probes a scene consisting of multiple interacting idealized targets.  Each sensor is a collocated transmit-receive pair that is capable of transmitting arbitrary wideband waveforms.  
We address the problem of finding the space-time transmit waveform that provides the best target detection performance in the sense of maximizing the energy scattered back into the receivers.  Our approach is based on earlier work that constructed the solution by an iterative time-reversal (TR) process.  In particular, for the case of idealized point-like scatterers in free space, we examine the frequency dependence of the eigenvalues of the TR operator, and we show that their behavior depends on constructive and destructive interference of the waves traveling along different paths.  
In addition, we show how these eigenvalues are connected to the poles of the Singularity Expansion Method.  
Our study of the frequency behavior distinguishes this work from most previous TR work, which focused on single-frequency waveforms and noninteracting targets.  
The main result of the present paper is that the TR process provides an automated, guaranteed way to find the resonant frequencies of the scattering system.
  
\end{abstract}

\begin{IEEEkeywords}
Time Reversal, DORT, eigenvalues, inverse scattering, multiple sensors, SEM (singularity expansion method), MIMO (multiple input multiple output), detection theory, Power Method
\end{IEEEkeywords}}

%% file: sections/body.tex
\section{Introduction}
% Computer Society journal papers do something a tad strange with the very
% first section heading (almost always called "Introduction"). They place it
% ABOVE the main text! IEEEtran.cls currently does not do this for you.
% However, You can achieve this effect by making LaTeX jump through some
% hoops via something like:
%
%\ifCLASSOPTIONcompsoc
%  \noindent\raisebox{2\baselineskip}[0pt][0pt]%
%  {\parbox{\columnwidth}{\section{Introduction}\label{sec:introduction}%
%  \global\everypar=\everypar}}%
%  \vspace{-1\baselineskip}\vspace{-\parskip}\par
%\else
%  \section{Introduction}\label{sec:introduction}\par
%\fi
%
% Admittedly, this is a hack and may well be fragile, but seems to do the
% trick for me. Note the need to keep any \label that may be used right
% after \section in the above as the hack puts \section within a raised box.


\IEEEPARstart{T}{his} paper addresses the problem of finding the space-time waveform that results in the most energy scattered back to the sensors from a distant target.  We consider the case in which a small number of ideal discrete sensors both transmit and receive, so that the problem becomes one of determining the time-domain waveforms that should be transmitted by the different sensors so that those sensors receive the maximum scattered energy.  

The problem of increasing the scattered energy received by the sensors has been previously studied by a number of authors.  
For a single sensor, \cite{hansen} proposed an approach based on trying dfferent frequencies in a random sequence.  
For multiple sensors, received signal energy can be increased by a process known as Time Reversal (TR) or  D\'ecomposition de l'Op\'erateur de Retournement Temporel (DORT),  in which scattered signals are recorded at the receivers and then time-reversed and retransmitted.
The TR idea, which has its roots in optical phase conjugation, has been used in numerous studies in acoustics and electromagnetism \cite{fink_basicprinciples,fink_TRM,dyab,rosny_TRM_EM}, and has potential applications  ranging from  sensing (remote imaging, radar, acoustics, optics, etc.) to communications  \cite{dyab,bogert,tortel,fink_TRM}.  
  
The lynchpin of the single-frequency TR approach is the application of the classical mathematical theorem known as the Power Method, which guarantees that under mild conditions, the application of successive powers of an operator to an arbitrary seed vector results in a sequence of vectors that converges to the eigenvector associated with the maximum eigenvalue \cite{isaacson_keller}.  In the sensing context, the relevant operator is the TR operator, which is the product of the scattering operator and its adjoint.  
The ``largest eigenvector" (i.e., the eigenvector corresponding to the largest eigenvalue) corresponds to specifying the relative strengths and phases of the transmitted fields in a way that causes the most energy to scatter back to the receivers.   This eigenvector contains information about the scatterers:  \cite{fink_convergence,prada_eigenmodes} showed that for point-like scatterers, the temporal wavefield focuses on the strongest scatterer \cite{fink_tracoustic,blomgren}; for a single spherical target,  \cite{chambers_singler_scatterer} showed that this eigenvector contains information about the target size and composition.  
The work \cite{jin_TR_antenna_arrays,moura_detection_single_antenna} showed that use of TR can improve target detection.  
 These analyses were all carried out for waveforms of a single temporal frequency under the single-scattering assumption (Born approximation).

Analysis of the TR temporal behavior for the general case of multiple scattering was investigated by \cite{cheneyoptimal,cheney_electromagnetics}, through an analysis of the frequency dependence of the TR operator.  This work showed that in general the time-domain TR construction converges automatically to a certain single-frequency waveform.  The frequency of this limiting waveform is a resonance, namely a frequency at which the largest eigenvalue has a local maximum, and the waveform's spatial shape (which determines its radiation pattern) is given by the corresponding eigenvector.  Thus the time-domain TR process can be used to tune automatically to the space-time waveform that is optimal in the sense of providing the best target detection performance. 
These predictions of Cheney {\it et al.} \cite{cheneyoptimal,cheney_electromagnetics} are consistent with experimental observations reported by \cite{kyritsi_channelcorrelation,fink_TRM,fink_convergence,fink_tracoustic}, who noticed that as the iterative TR process proceeds, the pulse temporally broadens, the frequency can shift, and the spectrum narrows.
The predictions are also consistent with the simulations of \cite{cherkaeva}, who also found that the iterative
TR process converges to a time-harmonic waveform.  

This paper carries out a detailed analysis of the case of two sensors and two point-like scatterers, where the scatterers are ``interacting" in the sense that there is multiple scattering between them.   
In particular, we obtain formulas for the scattering operator and the associated resonances; we show that these resonances arise from constructive interference between the various scattered waves.  
We show, moreover, that the TR method provides a stable experimental method for finding these resonances.

Our explicit consideration of multiple scattering is one aspect that differentiates this work from previous TR analysis.  

Another important connection between the present paper and earlier work is its relation to the Singularity Expansion Method (SEM) \cite{baum88}.  Specifically, the poles of the SEM are associated with information derived from the iterative TR process. The analysis of the eigenvalues associated with the TR matrix leads to an understanding of the relationship between the poles of the  SEM and the eigenvalues of the TR matrix.  In particular, 
%solving the system of equations derived from the Lippman-Schwinger formulation of the wave equation, 
the poles of the SEM appear in the expression for the scattered field.   
Here our approach uses Fourier analysis to avoid issues with the two-sided Laplace transform in existing SEM analyses.

This paper is organized as follows.  In Section \ref{formulation}, we provide some mathematical background on scattering theory.  In Section \ref{twopairs}, we specialize to the case of two point scatterers and two transmit/receive pairs, where  the Foldy-Lax method \cite{foldy1,foldy2} is used to obtain the scattering operator.  In Section \ref{scatteringoperator}, we obtain explicit analytical expressions for the scattering operator and its spectral decomposition.  In Section \ref{trprocess}, we outline the TR process and discuss its behavior in the two-target, two-sensor case.  In Section \ref{connectSEM}, we discuss the connection to the SEM poles. 
In Section \ref{eigenvaluesstability}, we show that the TR process and the eigenvalue information obtained from it are stable with respect to noise.  
 In Section \ref{numericalresults}, we show simulation results.    
 Section \ref{conclusion} gives conclusions and suggestions for future work.

%%%%%%%%%%%%%%%%%%%%%%%%%%%%%%
\section{Mathematical Formulation and Derivation}
\label{formulation}
We begin by solving the simplified scalar model of the wave equation for the time-domain electric field $E$,
\begin{equation}
\label{wave_equation}
\left( \nabla^2 - \frac{1}{c^2(\mathbf{r})} \frac{\partial^2}{\partial t^2}\right) E(\mathbf{r},t) =0,
\end{equation} where we think of $c(\mathbf{r})$ as the local speed of propagation of the electromagnetic (EM) waves in a specified media.  In free space, the speed is $c_0$.  
Scattering can be thought of as being produced by changes in the wave speed with perturbation (scattering potential)
\begin{equation}
\label{scattering_potential}
V(\mathbf{r}) =\frac{1}{c_0^2}  - \frac{1}{c^2(\mathbf{r})}.
\end{equation}  Under the Fourier Transform pair 
\begin{equation}
\label{forwardtransform}
\tilde{E}(\mathbf{r},\omega) = \int_{-\infty}^{\infty} E(\mathbf{r},t) e^{- j \omega t} \ud t,
\end{equation}

\begin{equation}	\label{inversetransform}
E (\mathbf{r},t) = \frac{1}{2 \pi	} \int_{-\infty}^{\infty} \tilde{E}(\mathbf{r},\omega) e^{j \omega t} \ud \omega,
\end{equation} (\ref{forwardtransform}) converts (\ref{wave_equation}) into 
\begin{equation}
(\nabla^2 + k^2- \omega^2 V(\mathbf{r}))\tilde{E}=0,
\end{equation} where $k = \omega/c_0$.  The convention throughout is that a tilde denotes a frequency-domain quantity.

The relevant free-space fundamental solution $\tilde{G}$, which satisfies
\begin{equation}
(\nabla^2 + k^2)\tilde{G} = -\delta(\mathbf{r}),
\end{equation}
 is the Green's function 
\begin{equation}
\label{green}
\tilde{G}(\mathbf{r},\omega) =- \frac{e^{-j k \|\mathbf{r}\|}}{4 \pi \| \mathbf{r} \|},
\end{equation} where $\|\cdot\|$ is the Euclidean norm, and $\delta$ is the Dirac delta. 

For a radiation source
\begin{equation}
\label{source}
\tilde{F}(\mathbf{r},\omega) = \mu_0 j \omega \tilde{J}(\mathbf{r},\omega),
\end{equation} where $\mu_0$ is the permeability of free space and $\tilde{J}$ is a scalar current density, the solution to the inhomogeneous wave equation  
\begin{equation}
\label{wave}
\left(\nabla^2 + k^2 \right) \tilde{E}_{\mathrm{in}}(\mathbf{r},\omega) = \tilde{F}(\mathbf{r},\omega)
\end{equation} is the radiated field
\begin{equation}
\label{frequencysolution}
\tilde{E}_{\mathrm{in}}(\mathbf{r},\omega) =  \int\limits_{\Omega} \tilde{G}(\mathbf{r}-\mathbf{y},\omega) \tilde{F}(\mathbf{y},\omega) \ud \mathbf{y},
\end{equation}  where the volume $\Omega$ of the source is a subset of $\mathbb{R}^3$ and $d \mathbf{y}$ denotes the volume element.  

The total field $\tilde{E}_{\mathrm{tot}}$ due to the source in the presence of scatterers satisfies
\begin{equation}
\label{total_field}
\left(\nabla^2 +k^2 -\omega^2 V(\mathbf{r}) \right) \tilde{E}_{\mathrm{tot}} (\mathbf{r}, \omega) = \tilde{F}(\mathbf{r},\omega).
\end{equation}
We can write the total field as a sum of the incident field in (\ref{frequencysolution}) plus the scattered field generated by a target, $\tilde{E}_{\mathrm{tot}} = \tilde{E}_{\mathrm{in}}+\tilde{E}_{\mathrm{sc}}$ \cite{cheney2009fundamentals}.  
  Subtracting (\ref{wave}) from (\ref{total_field}) yields
\begin{equation}
\label{sc}
\left(\nabla^2 + k^2 \right) \tilde{E}_{\mathrm{sc}}(\mathbf{r}, \omega) =-V(\mathbf{r}) \omega^2 \tilde{E}_{\mathrm{tot}} (\mathbf{r},\omega).  
\end{equation}
By convolving both sides of (\ref{sc}) with $\tilde{G}$, we obtain the \emph{Lippmann-Schwinger} equation for the scattered field at an arbitrary observation point $\mathbf{r}$
\begin{equation}
\label{scattered_field_frequency}
\tilde{E}_{\mathrm{sc}}(\mathbf{r},\omega) =  \int\limits_{\Omega} \tilde{G}(\mathbf{r}-\mathbf{y},\omega) \omega^2 V(\mathbf{y})\tilde{E}_{\mathrm{tot}}(\mathbf{y},\omega) \ud \mathbf{y}.
\end{equation} 

Equation (\ref{scattered_field_frequency}) is a Fredholm integral equation of the second kind \cite[pp.~98-101]{BarrySimon} and can be solved by a variety of different methods.  This Lippmann-Schwinger equation provides a framework for constructing the map from the source $\tilde{F}$ to the fields at observation point $\mathbf{r}$.  We call this map the scattering operator $S$.

In the next section, we work out the explicit special case of a scattering operator for 2 point scatterers and 2 transmit/receive pairs.

%%%%%%%%%%%%%%%%%%%%%%%%%%%%
\section{Scattering Operator for two point scatterers and two transmit/receive pairs}
\label{twopairs}
In this section, we examine the scattering operator $S$ for the special case of two point scatterers and two transmit/receive pairs of sensors. First the incident, scattered, and total fields are derived using the Foldy-Lax method \cite{devaney,foldy1,foldy2}. Then $S$ is obtained and recast in terms of three physically interpretable matrices: the forward and back-scattered propagators associated with the paths between sensors and scatterers; and an interaction matrix that characterizes the interactions among the scatterers.

The wave speed perturbation $V$ in (\ref{scattering_potential}) can be modeled as a sum of point scatterers.  For two point scatterers,  located at $\mathbf{x}_1$ and $\mathbf{x}_2$, (\ref{scattering_potential}) becomes
%Based on the Foldy-Lax formulation \cite{devaney,foldy1,foldy2}, the scattering potential $V$ from (\ref{scattering_potential}) can be a sum of idealized point scatterers
\begin{equation}
\label{potential}
V(\mathbf{r}) = \sum_{m=1}^{2} q_m  \delta (\mathbf{r} - \mathbf{x}_m),
\end{equation}
where $q_m$ is the scattering strength or the reflectivity  of the $m$th target. 

%%%%%%%%%%%%%%%%%%%%%%%%
\subsection{Foldy-Lax Method}
We use the Foldy-Lax method \cite{devaney,foldy1,foldy2} to obtain an explicit expression for the scattering operator.  
The Foldy-Lax method requires solving the set of linear equations ($n=1,2$):
\begin{align}
\label{localfields}
\tilde{E}^n (\mathbf{x}_n,\omega) &= \tilde{E}_{\mathrm{in}} (\mathbf{x}_n,  \omega) \nonumber \\ 
&+ \sum_{m\neq n} \tilde{G}( \mathbf{x}_n - \mathbf{x}_m,  \omega)  \omega^2 q_m \tilde{E}^m (\mathbf{x}_m,\omega).
\end{align}  
Here $\tilde{E}^m$ denotes the ``locally incident field'' at the scatterer $m$, namely the incident field seen by the $m$th scatterer.  Equation (\ref{localfields}) expresses the incident field at the scattering location $\mathbf{x}_n$ as a sum of the overall incident field $\tilde{E}_{\mathrm{in}}$ plus the sum of the scattered fields from all of the other scatterers. 

\begin{figure*}[!t]
\centering
\includegraphics[width=0.8\textwidth]{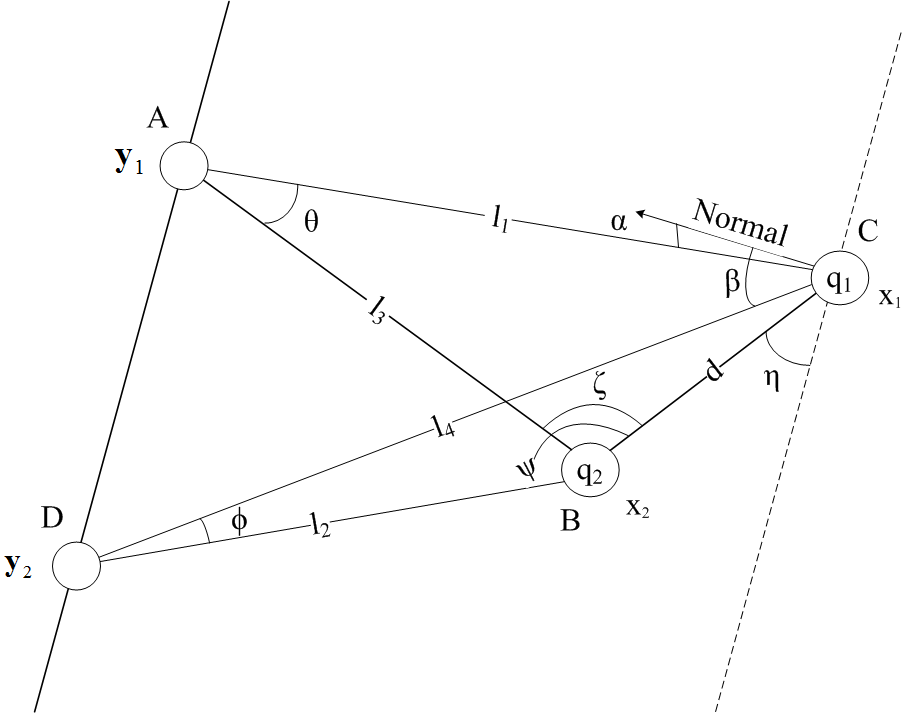}
\caption{Scattering geometry.  $\mathbf{y}_i$ denotes the transmit/receive pairs at points $C$ and $D$.  $l_i$ denotes the distances from the $\mathbf{y}_i$ to the scatterer at $\mathbf{x}_k$ at Points $A$ and $B$ with reflectivity coefficient $\mu_k$.}  
\label{scatteredfieldgeometry}
\end{figure*}

In \eqref{scattered_field_frequency} we use \eqref{potential} and substitute the ``locally incident field" of \eqref{localfields} for the ``total" field $\tilde{E}_{\mathrm{tot}}$ at the appropriate scatterer.  
The scattered field is then 
\begin{equation}
\label{scattered_field_foldy}
\tilde{E}_{\mathrm{sc}}(\mathbf{r},\omega) = \sum_{m=1}^{2} \tilde{G} (\mathbf{r}-\mathbf{x}_m , \omega ) \omega^2 q_m \tilde{E}^m (\mathbf{x}_m,\omega).
\end{equation}  
%Equation (\ref{scattered_field_foldy}) expresses the overall scattered field at some observation point $\mathbf{r}$ as a sum of scattered fields from each scatterer.  
Figure \ref{scatteredfieldgeometry} shows the possible direct paths of the fields to the point scatterers and their scattered returns.
   
%\begin{figure}[!h]
%\centering
%\includegraphics[width=3.5in]{./figures/farfield_geometry.png}
%\caption{Farfield approximation of Figure \ref{scatteredfieldgeometry} using $X_1$ as the reference scatterer. As the distance of the sensors and scatterers increases, $l_1 \approx l_4 \approx l$ and $l_2 \approx l_3$ in Figure \ref{scatteredfieldgeometry} result in the bisection of subtending angle by the segment connecting the midsection between the sensors to the reference scatterer.}  
%\label{farfieldgeometry}
%\end{figure} 

We re-write (\ref{localfields}) in matrix form
\begin{align}
\label{system-of-equation}
& \left(\begin{array}{cc}
1 & - \omega^2 q_2 \tilde{G}(\mathbf{d}, \omega) \\
- \omega^2 q_1 \tilde{G}(\mathbf{d},\omega) & 1
\end{array}\right)   \nonumber\\ \times & \left(\begin{array}{c}
\tilde{E}^1(\mathbf{x}_1,\omega) \\ \tilde{E}^2(\mathbf{x}_2,\omega)
\end{array}\right) = \left(\begin{array}{c}
\tilde{E}_{\mathrm{in}}(\mathbf{x}_1,\omega) \\
\tilde{E}_{\mathrm{in}}(\mathbf{x}_2,\omega)
\end{array}\right),
\end{align}where $\mathbf{d} =\mathbf{x}_1 - \mathbf{x}_2$.  The solution  to (\ref{system-of-equation}) is
\begin{eqnarray}
\label{totalfieldsolution}
\tilde{E}^1 (\mathbf{x}_1,\omega)= \frac{\tilde{E}_{\mathrm{in}}(\mathbf{x}_1,\omega) + q_{2} \omega^2 \tilde{G}(\mathbf{d}) \tilde{E}_{\mathrm{in}}(\mathbf{x}_{2}, \omega)}{1- \omega^4 q_1 q_2 \tilde{G}^2(\mathbf{d})} \\
\label{totalfieldsolution2}
\tilde{E}^2 (\mathbf{x}_2,\omega)=\frac{\tilde{E}_{\mathrm{in}}(\mathbf{x}_2,\omega) + q_1 \omega^2 \tilde{G}(\mathbf{d}) \tilde{E}_{\mathrm{in}}(\mathbf{x}_{1},\omega)}{1- \omega^4 q_1 q_2 \tilde{G}^2(\mathbf{d})} 
\end{eqnarray}
where for notational convenience we have suppressed the $\omega$ argument in $\tilde{G}$.  
Substituting (\ref{totalfieldsolution}) and (\ref{totalfieldsolution2}) into (\ref{scattered_field_foldy}) implies 
\begin{align}
\label{final_scattered_field}
\tilde{E}_{\mathrm{sc}}(\mathbf{r},\omega) = & \omega^2 q_1   \tilde{G}(\mathbf{r}-\mathbf{x}_1) \nonumber \\ \times & \frac{\tilde{E}_{\mathrm{in}}(\mathbf{x}_1,\omega) + \omega^2 q_2 \tilde{G}(\mathbf{d}) \tilde{E}_{\mathrm{in}}(\mathbf{x}_{2},\omega)}{1- \omega^4 q_1 q_2 \tilde{G}^2(\mathbf{d})}  \nonumber \\
& +\omega^2 q_2   \tilde{G}(\mathbf{r}-\mathbf{x}_2) \nonumber \\ \times & \frac{\tilde{E}_{\mathrm{in}}(\mathbf{x}_2,\omega) + \omega^2 q_1 \tilde{G}(\mathbf{d}) \tilde{E}_{\mathrm{in}}(\mathbf{x}_{1},\omega)}{1- \omega^4 q_1 q_2 \tilde{G}^2(\mathbf{d})}.
\end{align} 
Note that when $\omega^4 q_1 q_2 \tilde{G}^2(\mathbf{d}) < 1$, the denominator of \eqref{final_scattered_field} can be expanded in a geometric series, and in the resulting expression, each term can be interpreted in terms of scattering and propagation between sensors and scatterers.  This interpretation is discussed in more detail in Section  \ref{FrequencyDependence}.  

%%%%%%%%%%%%%%%%%%%%%%%%%%%%%
\subsection{Incident field and scattering matrix}

In this example, the source in (\ref{source}) is the sum of point-like sources from the transmitter locations $\mathbf{y}_i$

\begin{equation}
\tilde{F}(\mathbf{r},\omega) =\sum\limits_{i=1}^{2} \mu_0 j \omega \tilde{J}_i(\omega) \delta (\mathbf{r} - \mathbf{y}_i). \footnote{The units of $\tilde{J}$ are ampere-meters, which is due to the point-like volume source for our model.}
\end{equation} 
Consequently, the incident field in (\ref{frequencysolution}) is
\begin{align}
\label{totalincidentfield}
\tilde{E}_{\mathrm{in}} (\mathbf{r},\omega)  =& \sum\limits_{i=1}^{2} -\mu_0 j \omega \tilde{J}_i(\omega) \frac{e^{-j k \| \mathbf{r} - \mathbf{y}_i \|}}{4 \pi \| \mathbf{r} - \mathbf{y}_i \|}.
\end{align} 
Next we substitute (\ref{totalincidentfield}) into (\ref{final_scattered_field}) and evaluate the scattered field $\tilde{E}_{\mathrm{sc}}(\mathbf{r},\omega)$ at the sensor location $\mathbf{r}=\mathbf{y}_i$ to obtain the received signal for the $i$th sensor.  We denote this received field as 
\begin{align}
\label{receivedsignals}
& \tilde{R}_i(\omega) =\tilde{E}_{\mathrm{sc}}(\mathbf{y}_i,\omega) \nonumber \\
 & = \sum_{k=1}^{2} \frac{ \mu_0 j \omega \tilde{J}_k (\omega)}{\xi} \sum_{n=1}^{2}\tilde{G}(\mathbf{y}_i - \mathbf{x}_n,\omega) q_n \omega^2 \times   \nonumber \\ &  [\tilde{G}(\mathbf{x}_n - \mathbf{y}_k,\omega) 
 + q_{(n \; \mathrm{mod} \; 2) +1} \omega^2 \tilde{G}(\vb{d},\omega) \nonumber \\ 
 & \times \tilde{G}( \mathbf{x}_{(n \; \mathrm{mod} \; 2) + 1} - \mathbf{y}_k,\omega)],
\end{align} where $ (n\; \mathrm{mod}\; 2)+1$ is equal to 1 when $n=2$ and 2 when $n=1$, and
\begin{equation}		\label{xidef}
\xi = 1 - q_1 q_2 \omega^4 \tilde{G}^2(\vb{d},\omega).
\end{equation} The parameter $\xi$ is the determinant of the matrix in (\ref{system-of-equation}) and describes how the fields scatter between the two point scatterers.  If the scattering between the two scatterers is weak,  we can assume $\xi \approx 1$. 
 It turns out that $\xi$ is an important parameter that contributes key information about the resonant frequencies of the scattering system, a fact that is discussed in more detail below.

In  (\ref{receivedsignals}), define the inner summand as 
\begin{align}
\label{scatteredmatrix}
S_{ik} = &\frac{1}{\xi} \sum_{n=1}^{2}\tilde{G}(\mathbf{y}_i - \mathbf{x}_n ,\omega) \omega^2 q_n \nonumber \\
  \times & [\tilde{G}(\mathbf{x}_n - \mathbf{y}_k,\omega) + \omega^2 q_{(n \; \mathrm{mod} \; 2) +1} \tilde{G}(\vb{d},\omega) \nonumber \\
  \times & \tilde{G}( \mathbf{x}_{(n \; \mathrm{mod} \; 2) +1} - \mathbf{y}_k,\omega)].
\end{align}  The square matrix $(S_{ik})$ is the standard scattering operator $S$, also known as the multistatic data matrix or the transfer matrix. The matrix $S$ maps the transmitted signal vector 
\begin{equation}
\label{signalvector}
T(\omega) = \left[\begin{array}{c}
\mu_0 j \omega \tilde{J}_1 (\omega)\\
\mu_0 j \omega \tilde{J}_2 (\omega)
\end{array}\right],
\end{equation} to the received signal vector \footnote{Signals in (\ref{signalvector}) and (\ref{receivedvector}) are measured in volts.}
\begin{equation}
\label{receivedvector}
R(\omega) = \left[\begin{array}{c}
\tilde{R}_1 (\omega)\\
\tilde{R}_2 (\omega)
\end{array}\right].
\end{equation}
The components of these signal vectors are the transmitted and induced current densities at sensors 1 and 2, respectively.

%%%%%%%%%%%%%%%%%%%%%%%%%%%%%%%
\subsection{Analysis of the scattering matrix}
\label{scatteringoperator}

On careful examination, (\ref{scatteredmatrix}) can be expressed as 
\begin{equation}
\label{scatteringmatrixshortform}
S=  \frac{\omega^2}{\xi} P \Delta P^T,
\end{equation} the product of two propagator matrices ($P$ and $P^T$) and an interaction matrix $(\Delta)$,  where

\begin{equation}
\label{deltamatrix}
\Delta = \left[\begin{array}{cc}
q_1 & q_1 q_2 \omega^2 \tilde{G}(\vb{d},\omega) \\ 
q_1 q_2 \omega^2 \tilde{G}(\vb{d},\omega) & q_2
\end{array}\right],
\end{equation}

\begin{equation}
\label{propagatorback}
P = \left[\begin{array}{cc}
\tilde{G}(\mathbf{y}_1-\mathbf{x}_1, \omega) & \tilde{G}(\mathbf{y}_1- \mathbf{x}_2,\omega) \\
\tilde{G}(\mathbf{y}_2- \mathbf{x}_1,\omega) & \tilde{G}(\mathbf{y}_2-\mathbf{x}_2,\omega)
\end{array}\right],
\end{equation} and the superscript $T$ denotes matrix transposition. All three matrices are physically interpretable. In particular,  P is an operator that maps the fields at the scatterer locations $\mathbf{x}_1$ and $\mathbf{x}_2$ to the corresponding fields at receiver locations $\mathbf{y}_1$ and $\mathbf{y}_2$. The transpose $P^T$ maps the fields from the transmitters located at $\mathbf{y}_1$ and $\mathbf{y}_2$ to the fields at the scatterers.  The interaction matrix $\Delta$ describes the field interactions due to the scatterers.

The interaction matrix is singular at the frequencies $\omega_\alpha$ for which  
\begin{equation}
\label{singularity-xi}
\xi(\omega_\alpha) = 1 - q_1 q_2 \omega_\alpha^4 \tilde{G}^2(\vb{d},\omega_\alpha) = 0.  
\end{equation}   We note also that the interaction matrix $\Delta$ is aspect-independent; in other words, it does not depend on the spatial relationship between the targets and sensors.  Consequently, it could be useful for various applications such as target classification.  
Unfortunately, the interaction matrix \eqref{deltamatrix} cannot be measured directly; instead the scattering matrix \eqref{scatteringmatrixshortform} is the measurable quantity.   The scattering matrix $S$ depends on the viewing angles between sensors and targets because the propagator matrices  \eqref{propagatorback} do.  

One case in which the scattering matrix can be used to obtain information about the scatterers is when the coupling 
between the scatterers, namely $q_1 q_2 \omega^4 \tilde{G}^2(\vb{d},\omega)$, is weak.  
In this case, the off-diagonal elements of $\Delta$ are negligible, and $\xi =1$ from \eqref{xidef}.  
This case was addressed in \cite{prada,prada_eigenmodes}, which showed that with enough sensors, the $n$th eigenvalue of the scattering matrix corresponds to the $n$th strongest scatterer.

In this paper we consider the more general case in which there is coupling between the two scatterers; that is, we take into  account the effects of multiple scattering between the two scatterers.  In this case, the eigenvalues of the scattering operator do not necessarily correspond to individual scatterers.  To understand the connection between eigenvalues and the scattering strengths $q_i$, we carry out the following analysis.  
For notational convenience, we omit explicit references to frequency in the Green's function, eigenvalues, and the operators unless it is required.

%%%%%%%%%%%%%%%%%%%%%
\subsection{Spectral decomposition of interaction matrix $\Delta$}
The eigenvalues of $\Delta$ are
\begin{align}
\label{eigenvalues}
\lambda_{1,2} =\frac{(q_1 + q_2) \pm \sqrt{(q_1 - q_2)^2 + 4(q_1 q_2 \tilde{G}(\mathbf{d}) \omega^2)^2}}{2} 
\end{align} 
and the corresponding eigenvectors are
\begin{align}
X_1 &=\left[ \begin{array}{cc}- \frac{-q_1 +q_2  - \sqrt{(q_1 - q_2)^2 + 4 (q_1 q_2 \tilde{G}(\mathbf{d}) \omega^2)^2}}{2 \omega^2 \tilde{G}(\mathbf{d}) q_1 q_2 } & 1 \end{array} \right]^T\\
X_2 &= \left[\begin{array}{cc}
- \frac{-q_1 +q_2  + \sqrt{(q_1 - q_2)^2 + 4 (q_1 q_2 \tilde{G}(\mathbf{d}) \omega^2)^2}}{2 \omega^2 \tilde{G}(\mathbf{d}) q_1 q_2  } & 1
\end{array}\right]^T.
\end{align}
%At a given frequency $\omega$, for 
In the case that $q_1 > q_2$ and $\left|\frac{4 q_1 q_2 \tilde{G}(\mathbf{d}) \omega^2}{(q_1-u_2)^2}\right|<1$  (multiple scattering is weak), the eigenvalues can be written in a series by using  $\sqrt{1+x}= \sum\limits_{n=0}^{\infty}  \binom{1/2}{n} x^n$ for $|x|<1$ to obtain
\begin{align}
\label{eigen1}
\lambda_1 = q_1 +  \sum\limits_{n=1}^{\infty} \frac{q_1- q_2}{2} \binom{1/2}{n} \left(\frac{4 q_1 q_2 \tilde{G}(\mathbf{d}) \omega^2}{(q_1-q_2)^2}\right)^{n},\\
\label{eigen2}
\lambda_2 = q_2 -  \sum\limits_{n=1}^{\infty} \frac{q_1- q_2}{2}  \binom{1/2}{n} \left(\frac{4 q_1 q_2 \tilde{G}(\mathbf{d}) \omega^2}{(q_1-q_2)^2}\right)^{n}.
\end{align}  
When the higher-order terms in (\ref{eigen1}) and (\ref{eigen2}) are negligible, the eigenvalues reduce to
\begin{eqnarray}
\lambda_1 &\approx& q_1, \\
\lambda_2 &\approx&  q_2.
\end{eqnarray}  
%and $\xi=1$.  
Consequently, the eigenvalues of the interaction matrix $\Delta$ correspond directly to the scattering strengths when the multiple scattering between the scatterers is weak, which is the case analyzed in \cite{prada,prada_eigenmodes}.  

In the general case when the interaction between the scatterers can not be neglected, the interaction matrix can be decomposed into the form
\begin{equation}	\label{DeltaSpectralDecomp}
\Delta = X \Lambda X^{-1},
\end{equation} where $X$ is a matrix whose columns are the eigenvectors $\hat{X}_i=X_i/\|X_i\|$ for $i=1,2$, and $\Lambda$ is a diagonal matrix of the eigenvalues of $\Delta$.

%%%%%%%%%%%%%%%%
\subsection{Spectral decomposition of scattering matrix $S$} 

The spectral decomposition of $S$ can be written 
\begin{equation}
\label{eigendecompS}
S = Y \Gamma Y^{-1},
\end{equation} where $\Gamma$ is a diagonal matrix whose diagonal elements are the eigenvalues of $S$ and $Y$ is a matrix whose columns are the eigenvectors of $S$.  To relate the eigenvalues of $S$ to those of $\Delta$, we use \eqref{DeltaSpectralDecomp} in 
 (\ref{scatteringmatrixshortform}) to obtain 
\begin{equation}
\label{scatteringeigendecomp}
S = \frac{\omega^2}{\xi}(PX)\Lambda(X^{-1}P^T) =\frac{\omega^2}{\xi} P \Delta P^T.
\end{equation}  

We determine the eigenvalues $\gamma$ of $S$ by solving the characteristic equation
\begin{equation}
\label{characteristicA}
\det(S - \gamma I ) =0,
\end{equation} 
which, with  (\ref{scatteringeigendecomp}), becomes
\begin{align}
0 &=\det \left(\frac{\omega^2}{\xi} PX \Lambda X^{-1}P^T - (PX)\gamma I (PX)^{-1} \right) \nonumber\\
 &=    \det (PX) \det \left( \frac{\omega^2}{\xi} \Lambda X^{-1}P^T(PX) - \gamma I \right) \nonumber \\ &\hspace{4cm} \times  \det(PX)^{-1},
\end{align} which implies
\begin{equation}
\label{scatteringmatrixeigen}
\det\left( \frac{\omega^2}{\xi} \Lambda X^{-1}P^T PX - \gamma I \right) = 0.
\end{equation} Let
\begin{equation}
\label{Mmatrix}
M =  X^{-1}P^TPX .
\end{equation}
Then (\ref{scatteringmatrixeigen}) can be written as
\begin{equation}
\label{detM}
\det  \left( \frac{\omega^2}{\xi}\left[\begin{array}{cc}
\lambda_1 m_{11} & \lambda_1 m_{12} \\
\lambda_2 m_{21} & \lambda_2 m_{22}
\end{array}\right] - \gamma I\right) =0,
\end{equation} where $m_{ik}$ is $ik^\mathrm{th}$ element of matrix $M$ and is explicitly shown in Appendix \ref{appendMmatrix}.  Equation (\ref{detM}) implies that the eigenvalues $\{\gamma_1, \gamma_2\}$ of the scattering matrix $S$ are related to the eigenvalues $\{\lambda_1, \lambda_2 \}$ of the interaction matrix $\Delta$. Because the system is coupled, the contribution from each of the scatterers can be seen in each of the eigenvalues of $S$.  

Under certain propagation conditions, for example when the 
off-diagonal elements of $M$ are negligible, the eigenvalues $\gamma$ of $S$ in (\ref{detM}) can correspond directly to the eigenvalues $\lambda$ of $\Delta$. Consequently, the corresponding eigenvectors provide the phase information for each of the transmitters to transmit with the appropriate time delays on the signal such that the field energy will spatially focus at the point corresponding to $\lambda$.  In general, however, the relationship is more complicated.  The next section gives an experimental method, the Time-Reversal (TR) method, for obtaining information about the frequency dependence of the eigenvalues of $S$.  

%%%%%%%%%%%%%%%%%%%%%%%%%%%%%%%
\section{Time-Reversal Process}
\label{trprocess}
The TR process involves an  array of sensors (possibly distributed) that both transmits and receives.  The TR process is implemented via the following steps: 
\begin{enumerate}
\item Transmit a designated waveform (see below) from each transmitter.  Denote the signal transmitted from the $m$th transmitter by $T_m(t)$, so that the signal transmitted from the sensor array can be written $\mathbf T(t) =  [ T_1(t), T_2(t)]^T$.  

The transmitted field then scatters from the targets and propagates back to the sensors.
\item \label{receive} The sensors each receive the scattered signal and store it in some manner.  
Denote the signal received at the $n$th sensor by
\begin{equation}
\label{Rresponse}
R_n(t) = \sum\limits_{m=1}^{2} s_{nm}(t) * T_m(t),
\end{equation} 
where $*$ is convolution in time and $s_{nm}$ is the inverse Fourier transform of \eqref{scatteredmatrix}.  
\item These stored received signals are time reversed to obtain $\mathbf R(-t) = [ R_1(-t), R_2(-t)]^T$.
\item  The process repeats, with the next transmitted waveform being the previous time reversed-received signal: $\mathbf T(t) = \vb R(-t)$.  
\end{enumerate}
Since the process is straightforward to analyze in the frequency domain, where the matrix version of \eqref{Rresponse} is simply
\begin{equation}
\label{receivedsignal}
\mathbf{\tilde{R}}(\omega)= S(\omega) \mathbf{\tilde{T}}(\omega).
\end{equation} 
Here $S$ is a $2 \times 2 $ matrix and
\begin{equation}
\vb{\tilde{R}(\omega)} = \left[\begin{array}{c}
\tilde{R}_1(\omega) \\ \tilde{R}_2(\omega)
\end{array}\right], \quad \vb{\tilde{T}(\omega)} = \left[\begin{array}{c}
\tilde{T}_1(\omega) \\ \tilde{T}_2(\omega)
\end{array}\right].
\end{equation} 
Moreover, the time-reversed received signal vector is the phase conjugate of the received signal $\tilde{\vb{R}}$.

To illustrate the TR process, let  the first transmitted vector be
\begin{equation}		\label{initialsignal}
\mathbf{\tilde{T}}^0(\omega) = \left[\begin{array}{c}
\mu_0 j \omega \tilde{J}_1(\omega) \\ \mu_0 j \omega \tilde{J}_2(\omega)
\end{array}\right],
\end{equation} where $\tilde{J}_i$ is the current density induced on the $i^\mathrm{th}$ antenna.  The corresponding received vector is
\begin{equation}
\mathbf{\tilde{R}}^0 (\omega)=S\mathbf{\tilde{T}}^0(\omega).
\end{equation}  
After the received signal is phase conjugated, the new transmitted signal is
\begin{equation*}
\mathbf{\tilde{T}}^1(\omega) = \mathbf{\tilde{R}}^{0*} (\omega)= S^* \mathbf{\tilde{T}}^{0*}(\omega),
\end{equation*} where the asterisk in the exponent denotes complex conjugation.  Consequently, the next received signal is
\begin{equation*}
	\mathbf{\tilde{R}}^1(\omega) = S \mathbf{\tilde{T}}^1 (\omega) = S S^* \mathbf{\tilde{T}}^{0*}(\omega) .
\end{equation*} Therefore, the transmit signals at the second and the third iterations are 
\begin{eqnarray*}
\mathbf{\tilde{T}}^2(\omega) &=&  S^*S \mathbf{\tilde{T}}^0 (\omega)\\
\mathbf{\tilde{T}}^3(\omega) &=& S^*S S^* \mathbf{\tilde{T}}^{0*}(\omega).
\end{eqnarray*}  Generally for the even and odd iterations,
\begin{equation}
\begin{array}{lcl}
\mathbf{\tilde{T}}^{2n} (\omega) &=& \left(S^*S\right)^n \mathbf{\tilde{T}}^0 (\omega)\\
\label{oddsignals}
\mathbf{\tilde{T}}^{2n+1} (\omega) &=& \left(S^*S\right)^n S^* \mathbf{\tilde{T}}^{0*} (\omega).
\end{array}
\end{equation} 
These transmitted signals are identical to the ones derived by Prada and Fink \cite{fink_TRM,fink_basicprinciples}.   
% For the $(2n)^{\mathrm{th}}$ transmitted signal, the time reversed matrix is raised to the $n^{\mathrm{th}}$ power.  
The well-known Power Method \cite{isaacson_keller} guarantees that with any choice of initial non-zero vector \eqref{initialsignal} and appropriate normalization, in practice the TR process converges to the eigenvector of the TR matrix associated with the largest eigenvalue, provided that this largest eigenvalue is not degenerate.  If the largest eigenvalue is degenerate, the TR process still provides a vector in the eigenspace associated with the largest eigenvalue.  

The classical Power Method \cite{isaacson_keller} and the Prada-Fink analyses \cite{fink_TRM,fink_basicprinciples} apply to a single time-harmonic wave.  Cheney {\it et al.} \cite{cheneyoptimal,cheney_electromagnetics} showed that the time-domain iterative TR process converges in general to a single time-harmonic wave; in other words, the TR process sharpens peaks in the spectrum.  
 In this paper, we further analyze the behavior of the TR process as a function of frequency.  

%%%%%%%%%%%%%%%%%%%%
\subsection{Spectral decomposition of TR matrix $L$}
Since $S$ is symmetric, $S^\dag = S^*$.
We define the TR matrix to be
\begin{equation}
\label{time_reversal_matrix}
L = S^* S = S^\dagger S.
\end{equation} Because $L$ is Hermitian, its eigenvalues $\{\sigma_1, \sigma_2\}$ are real-valued, and its eigenvectors are orthonormal. In particular, if $\Sigma$ denotes the diagonal matrix consisting of the eigenvalues of $L$, and $V$ denotes the unitary matrix whose columns are the corresponding eigenvectors of $L$, then the eigendecomposition of the TR matrix is 
\begin{equation}
\label{TRdecomp}
L = V \Sigma V^\dagger.
\end{equation} 

%%%%%%%%%%%%%
\begin{figure}[!h]
\centering
\includegraphics[width=3.5in]{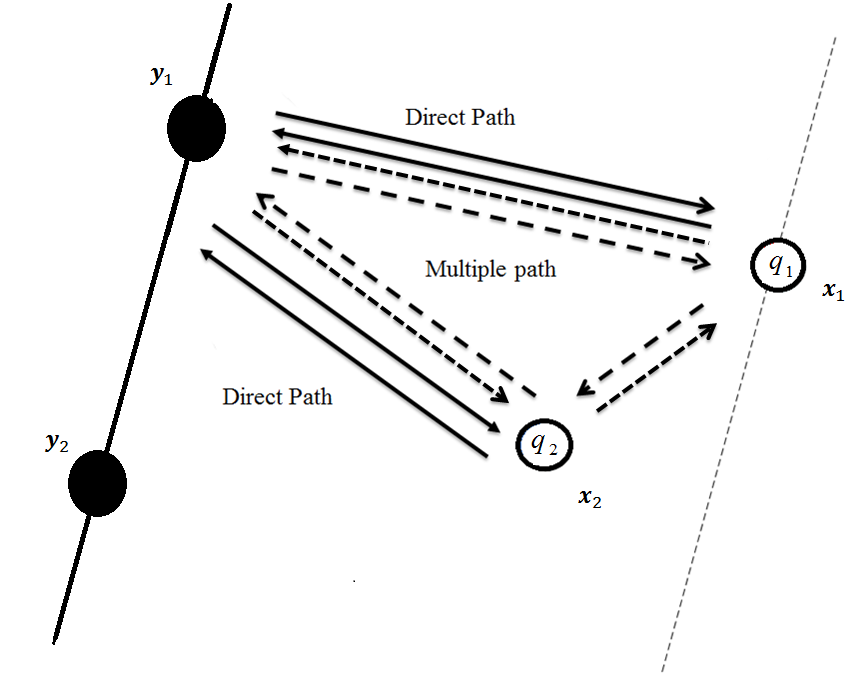}
\caption{Four propagation paths associated with transmitted signal from sensor $\vb{y}_1$ in  (\ref{a}).}  
\label{awave}
\end{figure}
%%%%%%%%%%%%%%

For this two-sensor problem,
\begin{equation}	\label{Sabc}
S = \begin{pmatrix} a & c \\
	c &  b
	\end{pmatrix}
\end{equation}
where
\begin{align}
\label{a}
a =  & \left(  \omega^2 \tilde{G}(\vb{y}_1 -\vb{x}_2)q_2 \tilde{G}(\vb{x}_2 - \vb{x}_1) q_1 \tilde{G} (\vb{x}_1 - \vb{y}_1)  \right. \nonumber \\ 
& +\omega^2 \tilde{G} (\vb{y}_1 -\vb{x}_1) q_1 \tilde{G}(\vb{x}_1 - \vb{x}_2) q_2 \tilde{G}(\vb{x}_2 - \vb{y}_1) \nonumber \\ 
& +\tilde{G}(\vb{y}_1 - \vb{x}_1) q_1 \tilde{G}(\vb{x}_1 - \vb{y}_1)    \nonumber \\ & \left. +\tilde{G}(\vb{y}_1 - \vb{x}_2) q_2 \tilde{G}(\vb{x}_2 - \vb{y}_1)  \right) \frac{\omega^2}{\xi}, \\
\label{b}
b = & \left( \omega^2 \tilde{G}(\vb{y}_2- \vb{x}_2) q_2 \tilde{G}(\vb{x}_2 - \vb{x}_1) q_1 \tilde{G}(\vb{x}_1 - \vb{y}_2) \right. \nonumber \\ &+ \omega^2 \tilde{G}(\vb{y}_2 - \vb{x}_1) q_1 \tilde{G}(\vb{x}_1 - \vb{x}_2) q_2 \tilde{G}(\vb{x}_2 - \vb{y}_2) \nonumber \\ &+ \tilde{G}(\vb{y}_2 - \vb{x}_2) q_2 \tilde{G}(\vb{x}_2 - \vb{y}_2)  \nonumber \\ & \left. + \tilde{G}(\vb{y}_2 - \vb{x}_1) q_1 \tilde{G} (\vb{x}_1 - \vb{y}_2) \right) \frac{\omega^2}{\xi},\\
\label{c}
c = &\left( \tilde{G}(\vb{y}_2 - \vb{x}_1) q_1 \omega^2 \tilde{G}(\vb{x}_1 - \vb{x}_2) q_2 \tilde{G} (\vb{x}_2 - \vb{y}_1) \right.   \nonumber \\ & + \tilde{G} (\vb{y}_2 - \vb{x}_2) q_2 \tilde{G}(\vb{x}_2 - \vb{y}_1) \nonumber \\ & + \tilde{G}(\vb{y}_2 - \vb{x}_2) q_2 \tilde{G} (\vb{x}_2 - \vb{x}_1) q_1 \omega^2 \tilde{G} (\vb{x}_1 - \vb{y}_1) \nonumber \\ & \left. + \tilde{G}(\vb{y}_2 - \vb{x}_1) q_1 \tilde{G}(\vb{x}_1 - \vb{y}_1) \right) \frac{\omega^2}{\xi} .
\end{align} 
%\begin{align}
%\label{a}
%a &= \left(2 q_1 q_2 \omega^2 \tilde{G}(\mathbf{d}) \tilde{G}(\mathbf{y}_1 - \mathbf{x}_1) \tilde{G}(\mathbf{y}_1- \mathbf{x}_2) \right. \nonumber \\ & \left.+ q_1 \tilde{G}^2(\mathbf{y}_1 - \mathbf{x}_1) + q_2 \tilde{G}^2(\mathbf{y}_1 - \mathbf{x}_2) \right) \frac{\omega^2}{\xi} \\
%\label{b}
%b &= \left(2 q_1 q_2 \omega^2 \tilde{G}(\mathbf{d}) \tilde{G} (\mathbf{y}_2- \mathbf{x}_1) \tilde{G}(\mathbf{y}_2 - \mathbf{x}_2) \right. \nonumber \\ & \left. + q_1 \tilde{G}^2(\mathbf{y}_2 - \mathbf{x}_1) + q_2 \tilde{G}^2(\mathbf{y}_2 - \mathbf{x}_2)\right) \frac{\omega^2}{\xi} \\
%\label{c}
%c &= \left( q_2 \tilde{G} (\mathbf{y}_1 - \mathbf{x}_2) \left[u_1  \omega^2 \tilde{G}(\mathbf{d}) \tilde{G}(\mathbf{y}_2 - \mathbf{x}_1) \right. \right. \nonumber \\ &+ \left. \tilde{G}(\mathbf{y}_2 - \mathbf{x}_2) \right] + q_1 \tilde{G}(\mathbf{y}_1 - \mathbf{x}_1) \left[u_2 \omega^2 \tilde{G}(\mathbf{d}) \right. \nonumber \\ & \left. \times \left.  \tilde{G}(\mathbf{y}_2 - \mathbf{x}_2)  + \tilde{\mathbf{G}}(\mathbf{y}_2 - \mathbf{x}_1)\right] \right) \frac{\omega^2}{\xi}.
%\end{align} 
The derivations of these expressions are provided in Appendix \ref{appendvalues}.
In terms of \eqref{a}-\eqref{c}, the trace of $L$ is %of (\ref{TReigens}) is 
\begin{equation}		\label{traceL}
\mathrm{tr} (L) = \left( |a|^2 + |b|^2 + 2 |c|^2 \right),
\end{equation} and the determinant of $L$ is
\begin{align}
\label{detL}
\det(L) &= |\det(S)|^2 =  |ab - c|^2.
%\left( |c|^4 + |a|^2 |b|^2 - 2 \mathrm{Re }\  a^*c^2 b^*  \right).
\end{align}   
The eigenvalues of a $2 \times 2$ matrix can be written in terms of the trace and determinant as
\begin{equation}
\label{TRsigma}
\sigma_{1,2}= {\mathrm{tr}(L) \pm \sqrt{\mathrm{tr}(L)^2 - 4 \det(L)} \over 2}.
\end{equation}  Because the eigenvalues are real valued, tr$(L)$ and det$(L)$ must be real, and in particular  
%Since the eigenvalues of $L$ are all real, 
the discriminant of (\ref{TRsigma}) cannot be negative.  Therefore, $4\det(L)/\mathrm{tr}(L)^2<1$.   We can expand the square root in (\ref{TRsigma}) and approximate both eigenvalues by their first-order terms 
\begin{align}
\label{largeeigenvalue}
\sigma_1 &\approx \mathrm{tr} (L) - \frac{\det(L)}{\mathrm{tr}(L)}\\
\label{smallereigenvalue}
\sigma_2 &\approx  {\det (L) \over \mathrm{tr} (L)}.
\end{align}

%\begin{figure}[!h]
%\centering
%	\subfloat[First interfering terms of $c$ in (\ref{c}) \label{cfirst}]{\includegraphics[width=.5\textwidth]{./figures/cwavepropfirst.png}}\\
%	\subfloat[Second interfering terms of $c$ in (\ref{c} \label{csecond})]{\includegraphics[width=0.5\textwidth]{./figures/cwavepropsecond.png}}
%	\label{subfigs}
%\end{figure}
%%%%%%%%%%%%%%%%%%%
\begin{figure}[!h]
\centering
\includegraphics[width=3.5in]{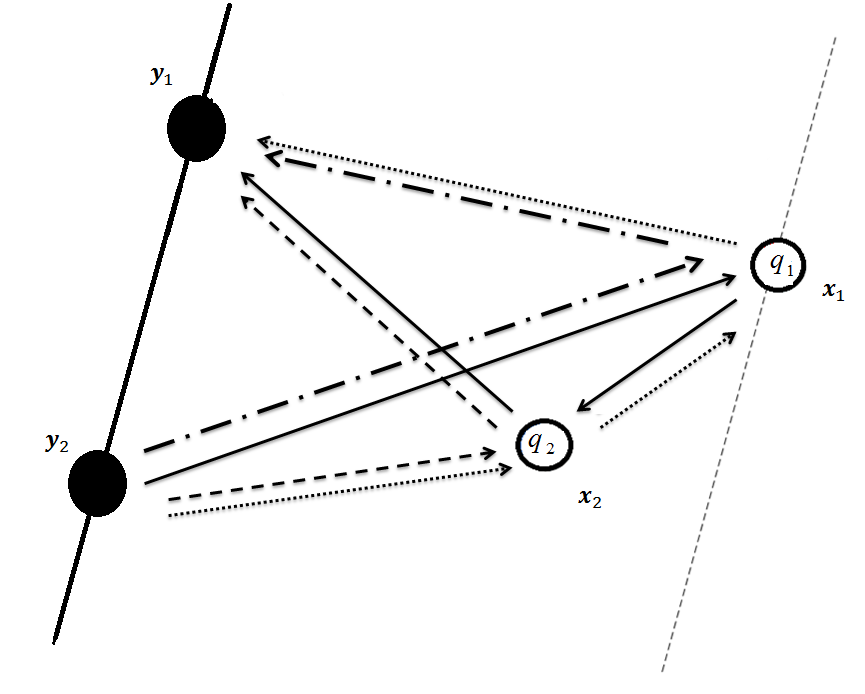}
\caption{Propagation paths of $c$ in (\ref{c}) from sensor $\vb{y}_2$.}  
\label{cfirst}
\end{figure}
%%%%%%%%%%%%%%%%%%%%
%\begin{figure}[!h]
%\centering
%\includegraphics[width=0.5\textwidth]{./figures/cwavepropsecond.png}
%\caption{Second interfering terms of $c$ in (\ref{c}).}  
%\label{csecond}
%\end{figure}
%%%%%%%%%%%%%%%%%%%%%%%

\subsection{Frequency dependence of  TR matrix $L$}		\label{FrequencyDependence}
The quantities $a$, $b$, and $c$ of \eqref{Sabc} have physical meaning:  they describe how the scatterers diffract the waves.  To understand their physical interpretations, we inspect the product of $S$ with the $k^\mathrm{th}$ iterate of the transmitted vector in the Power Method of (\ref{initialsignal})-(\ref{oddsignals}):
\begin{eqnarray}
\label{products}
S(\omega) \tilde{\vb{T}}^k(\omega) &=& \left( \begin{array}{cc} 
a & c \\ c & b
\end{array}\right)\left( \begin{array}{c} 
\tilde{T}^k_1 (\omega) \\ \tilde{T}^k_2(\omega)
\end{array}\right) \nonumber \\ &=&
\left(\begin{array}{c}
a \tilde{T}^k_1 (\omega) + c \tilde{T}^k_2 (\omega) \\ c \tilde{T}^k_1 (\omega) + b \tilde{T}^k_2 (\omega) 
\end{array} \right).
\end{eqnarray} Since the first component $a \tilde{T}^k_1 + c \tilde{T}^k_2$ is a sum of transmitted signals from sensor $\vb{y}_1$ $(a \tilde{T}^k_1)$ and sensor $\vb{y}_2$ $(c \tilde{T}^k_2)$, then $a$ and $c$ are associated with the propagation paths starting from sensors $\vb{y}_1$ and $\vb{y}_2$, respectively.  Similarly, for the second component $(c \tilde{T}^k_1 + b \tilde{T}^k_2)$, $c$ and $b$ are associated with propagation paths starting from sensor $\vb{y}_1$ $(c \tilde{T}^k_1)$ and sensor $\vb{y}_2$ $(b \tilde{T}^k_2)$, respectively.

 As depicted in Fig. \ref{awave}, the wave propagation corresponding to $a$ has four propagation paths associated with a signal that is transmitted from and received by sensor $\vb{y}_1$.  The first term in $a$ describes a wave propagating from the sensor $\mathbf{y}_1$  to the scatterer at $\mathbf{x}_1$ where it scatters with strength $q_1$,  then travels to  $\mathbf{x}_2$ where it is scattered with strength $q_2$, and finally propagates from $\mathbf{x}_2$ to $\mathbf{y}_1$ $(\text{path } \vb{y}_1 \to \vb{x}_1 \to \vb{x}_2 \to \vb{y}_1)$.  Similarly, the second term corresponds to the path $\vb{y}_1 \to \vb{x}_2 \to \vb{x}_1 \to \vb{y}_1$.
The third term of $a$ describes the direct-path scattering, in which the wave propagates from $\mathbf{y}_1$ to the scatterer at $\mathbf{x}_1$, where it is reflected with strength $q_1$ directly back to sensor $\mathbf{y}_1$ $(\vb{y}_1 \to \vb{x}_1 \to \vb{y}_1)$.  
The fourth term describes the corresponding direct-path scattering from the scatterer at $\mathbf{x}_2$ back to $\vb{y}_1$ $(\vb{y}_1 \to \vb{x}_2 \to \vb{y}_1)$.   The factor $\xi$ corresponds to waves reverberating between the scatterers.  

Similarly, expression $b$ represents the four paths emanating from and returning to sensor $\mathbf{y}_2$.  

Expression $c$ has two interpretations, depending on whether the signal starts from sensor $\vb{y}_1$ or sensor $\vb{y}_2$.  If the signal is transmitted from sensor $\vb{y}_2$ $(c \tilde{T}^k_2)$, the first term in (\ref{c}) represents propagation from sensor $\vb{y}_2$ to scatterer $\vb{x}_1$ to scatterer $\vb{x}_2$ and finally to sensor $\vb{y}_1$ $(\vb{y}_2 \to \vb{x}_1 \to \vb{x}_2 \to \vb{y}_1)$.  The second through fourth terms describe the remaining three paths $(\vb{y}_2 \to \vb{x}_2 \to \vb{y}_1,\  \vb{y}_2 \to \vb{x}_2 \to \vb{x}_1 \to \vb{y}_1, \  \vb{y}_2 \to \vb{x}_1 \to \vb{y}_1)$ in Fig. \ref{cfirst}.  Each path starts at sensor $\vb{y}_2$ and ends at sensor $\vb{y}_1$.  Now if the signal is transmitted by sensor $\vb{y}_1$ $(c \tilde{T}^k_1)$, the four terms in $c$ are interpreted as all possible single-scatter paths from sensor $\vb{y}_1$ to sensor $\vb{y}_2$: $\vb{y}_1 \to \vb{x}_2 \to \vb{x}_1 \to \vb{y}_2,\  \vb{y}_1 \to \vb{x}_2 \to \vb{y}_2,\  \vb{y}_1 \to \vb{x}_1 \to \vb{x}_2 \to \vb{y}_2 , \ \vb{y}_1 \to \vb{x}_1 \to \vb{y}_2 $.   We observe that these four paths are the reverse of the four paths when the signal emanates from sensor $\vb{y}_2$.

%Expression $c$ (See Figs. \ref{cfirst} and \ref{csecond}) describes the waves that begin at one sensor and end at the other.  The first term in $c$ (see Figure \ref{cfirst}) represents wave propagation from the sensor at $\mathbf{y}_2$ to the scatterer at $\mathbf{x}_1$, then to $\mathbf{x}_2$ and finally to the other sensor at $\mathbf{y}_1$.  The second term (see Figure \ref{csecond}) describes wave propagation from the source at $\mathbf{y}_2$ to a scatterer at $\mathbf{x}_2$ and then to the sensor at $\mathbf{y}_1$.  The third and fourth terms are similar except that the waves start at  $\mathbf{y}_1$ and end at $\mathbf{y}_2$ instead.

The largest eigenvalue of the TR operator $L$ in (\ref{largeeigenvalue}) represents the contribution of all the wave interactions that occur in the medium measured at the sensor locations.   
%When  $\det(L)/\mathrm{tr}(L)^2 <1$ is made, it turns out that the largest eigenvalue is approximately the trace of the TR matrix and represents the interferences of the wave propagation.  The diffraction maxima occurring at certain frequencies account for the peaks in the eigenvalue.  
The maximum value for the eigenvalue $\sigma_1$ of $L$  occurs when the waves described in $a$, $b$, and $c$ all constructively add.  This occurs when the path differences between every pair of propagation paths are equal to integer multiples of some fundamental wavelength $\lambda_0$.  From the many redundant equations corresponding to this constructive addition condition, only the following three relations are needed to specify that all waves add constructively.     
%%%%%%%%%%%%%
\begin{figure*}[!t]
\centering
\includegraphics[width=0.8\textwidth]{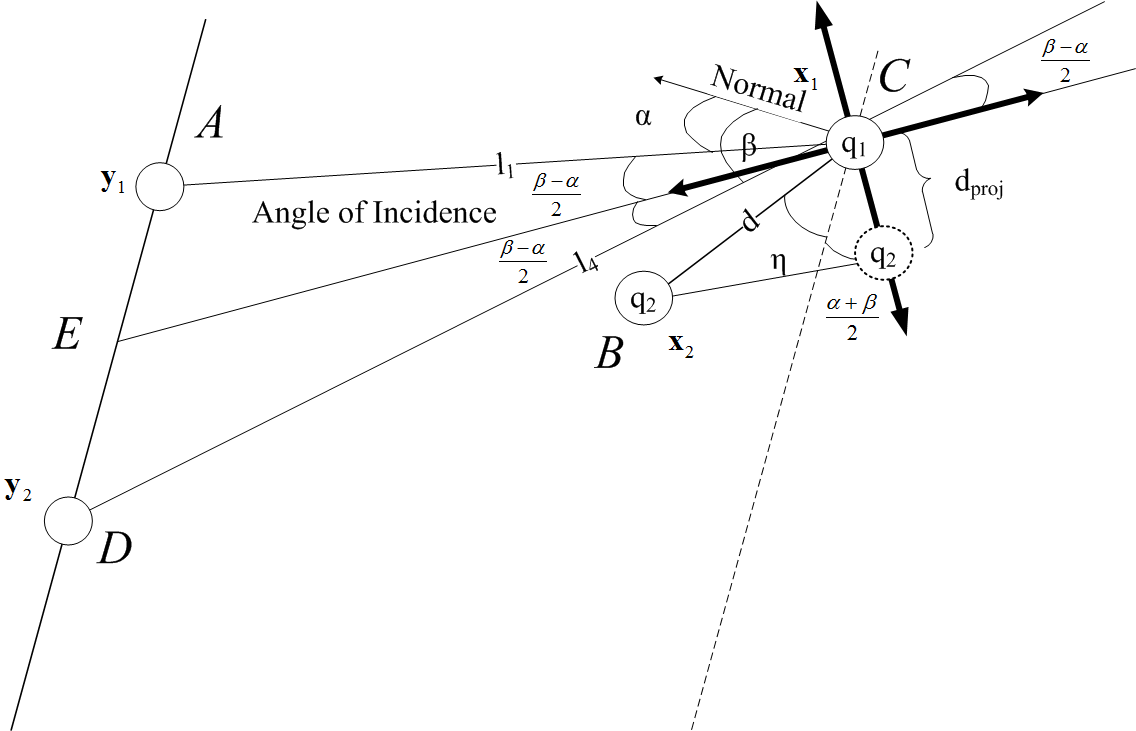}
\caption{Geometry of the distributed sensors and scatterers $d << l_k$, $l_k \to \infty$.  Farfield approximation of Figure \ref{scatteredfieldgeometry} using $\vb{x}_1$ as the reference scatterer. As the distance of the sensors and scatterers increases, $l_1 \approx l_4 \approx l$ and $l_2 \approx l_3$ in Figure \ref{scatteredfieldgeometry} result in the bisection of subtending angle by the segment connecting the midsection between the sensors to the reference scatterer. }  
\label{bragg_geometry}
\end{figure*} 
%%%%%%%%%%%%%%%%%%%
\begin{enumerate} 
\item From Figure \ref{scatteredfieldgeometry}, we see that
the difference between the direct-path scattering $\mathbf{y}_1 \longleftrightarrow \mathbf{x}_2$ and the path   $\mathbf{y}_1 \to \mathbf{x}_2 \to \mathbf{x}_1 \to \mathbf{y}_1$ must be an integer multiple of the wavelength $\lambda_0$:
\begin{equation}
\label{y1tox2}
{2 l_3 -(l_3 + d +l_1) \over \lambda_0} ={l_3 -d -l_1 \over \lambda_0}= m,
\end{equation} where $m$ is an integer.  
\item The path difference between the direct-path scattering $\mathbf{y}_2 \longleftrightarrow \mathbf{x}_1$ and the path $\mathbf{y}_2 \to \mathbf{x}_2 \to \mathbf{x}_1 \to \mathbf{y}_2$, must be an integer multiple of the wavelength:
\begin{equation}
\label{y2tox1}
{2 l_4 - (l_2+d+l_4) \over \lambda_0} = {l_4 -d-l_2 \over \lambda_0}=p,
\end{equation} where $p$ is an integer.
\item 
The path difference between paths $\mathbf{y_2} \to \mathbf{x}_2 \to \mathbf{y}_1$ and $\mathbf{y_2} \to \mathbf{x}_1 \to \mathbf{x}_2 \to \mathbf{y}_1$ must be an integer multiple of the wavelength:
\begin{equation}
\label{y2toy1}
{l_2+l_3 -(l_4+d+l_3) \over \lambda_0} = {l_2 - l_4 -d \over \lambda_0 } = r,
\end{equation} where $r$ is an integer.  
%\item Lastly, the path difference between paths $\mathrm{y}_1 \to \mathbf{x}_1 \to \mathbf{y}_2$ and $\mathbf{y}_1 \to \mathbf{x}_2 \to \mathbf{x}_1 \to \mathbf{y}_2$ must be an integer multiple of the wavelength,
%\begin{equation}
%\label{y1toy2}
%{l_4+l_1 - (l_3 + d + l_4) \over \lambda_0} = {l_1 -l_3 -d \over \lambda_0 } = s,
%%{l_4+l_1 - (l_2 + d + l_1) \over \lambda_0} = {l_4 -l_2 -d \over \lambda_0 } = s,
%\end{equation}where $s$ is an integer.  
\end{enumerate}

To obtain a single equation that involves the four propagation distances $\{l_1,l_2,l_3,l_4\}$, we add (\ref{y2toy1}) and (\ref{y2tox1}) to get $-2d/\lambda_0 =p+r$, and substitute this result into sum of  (\ref{y1tox2}) and (\ref{y2tox1}).  Consequently, 
\begin{equation}
\label{relation}
{l_3+l_4-l_1-l_2 \over \lambda_0} = m+ p -(p+r) =n,
\end{equation} where $n$ is an integer.  By the law of sines,  
\begin{eqnarray} \label{l1}
l_1 &=&  {d \sin \zeta \over \sin \theta}\\
\label{l2}
l_2 &=& {d \sin (\phi + \psi) \over \sin \phi}\\
\label{l3}
l_3 &=& {d \sin(\zeta+ \theta) \over \sin \theta}\\
\label{l4}
l_4 &=& {d \sin \psi \over \sin \phi},
\end{eqnarray} which represent each $l_k$ in terms of $d$ and various incident angles in Fig. 1.  After some manipulation (Appendix \ref{append:bragg}),
Equation (\ref{relation})  becomes
\begin{align}
\label{newrelation}
(l_4-l_2) + & (l_3-l_1) = d   \left( {\sin \psi - \sin(\phi+\psi) \over \sin \phi } \right. \nonumber \\ & + \left. {\sin(\zeta +\theta) - \sin \zeta \over \sin \theta }\right) = n \lambda_0,
\end{align}  which is  Bragg's condition for constructive interference \cite[pp.~706-707]{born}.  As the distances from the sensors to the scatterers approach infinity, $l_k \to \infty$, which implies that $\theta \to 0$, $\phi \to 0$, and $(\beta-\alpha)/2$ bisects the angle subtended by vectors from $\vb{x}_1$ to the two sensors.   
%.  Consequently, $\psi \to {\pi \over 2} + \eta + \beta$ and $\zeta \to {\pi \over 2} +\eta +\alpha $, where $\beta$ is the  incident angle from $\mathbf{y}_2$ to $\mathbf{x}_1$ and $\alpha$ is the incident angle from $\mathbf{y}_1$ to $\mathbf{x}_1$, and $\eta$ is the angle from one reference scatterer to another scatterer measured from the plane that includes the reference point and is parallel to the plane where the transmit/receive pairs are located in Figure \ref{scatteredfieldgeometry}  
In the far-field limit, \eqref{newrelation} becomes (see Appendix \ref{append:bragg} for the proof)
\begin{align}
\label{finalrelation}
n \lambda_0 &\approx d(-\cos \psi + \cos \zeta) \nonumber \\ &\approx d \left( \sin (\eta+\beta) - \sin (\eta+\alpha)\right) \nonumber \\ &= 2 d \cos \left( \eta + {\alpha + \beta \over 2}\right) \sin \left({\beta-\alpha \over 2 }\right).
\end{align}

Equation \eqref{finalrelation} can be understood in terms of Bragg's Law with the help of Fig. \ref{bragg_geometry}, where the four thick axis vectors in Fig. \ref{bragg_geometry} define a relevant reference frame. The angle between the axis vector from $C$ to the sensor midpoint $E$ and the ``normal" is $(\alpha + \beta)/2$.  This angle, when rotated by $90^{\circ}$, is the angle in Fig. \ref{bragg_geometry} labelled $(\alpha + \beta)/2$.  
Consequently, Bragg's Law is 
\begin{equation}
2 d_{\mathrm{proj}} \sin \left({\beta - \alpha \over 2} \right)
\approx n \lambda_0
\end{equation} 
where 
% Figure \ref{bragg_geometry} shows the geometry needed for the connection with Bragg's Law.     and
  $d_{\mathrm{proj}}=d \cos  \left(\eta + \frac{\alpha + \beta}{2} \right)$ is the projection of $d$ onto the line perpendicular to the bisecting vector. 
  
%  normal vector that has been rotated by $\eta + \frac{\alpha+\beta}{2}$.  These angles can be better seen in Figure \ref{bragg_geometry} where we rotated the axis of the reference scatterer in Figure \ref{scatteredfieldgeometry}.  $\frac{\beta - \alpha}{2}$ can be seen as the angle of incidence in the rotated axis.  The geometry in Figure \ref{bragg_geometry} can be further simplified to Figure \ref{farfieldgeometry} when the scatterers and the sensors are far from each other such that $l_1\approx l_4$, $l_2 \approx l_3$, and $d<<l_j$.  The line connecting from midpoint to the reference scatterer is the main axis where the normal can be drawn.  In Figure \ref{farfieldgeometry}, we can see that this line will approximately bisect the angle subtended by the lines connecting the sensors to the scatterer, and $d_\mathrm{proj}$ is the projected distance of $d$ onto the normal. 
% 
From the preceding analysis, Bragg's Law also gives the frequencies at which the largest eigenvalue $\sigma_1$ of the TR matrix $L$ achieves a maximum.  This largest $\sigma_1$ is the maximum field intensity at both sensors.
The smaller eigenvalue $\sigma_2$ in (\ref{smallereigenvalue}) is the ratio of the determinant and the trace of $L$. 
From (\ref{detL}), the determinant is  $|ab - c|^2$. The quantities $a$ and $b$ correspond to propagation paths beginning and ending at $\vb{y}_1$ and $\vb{y}_2$, respectively. Since $c$, which corresponds to propagation paths that begin at one sensor and end at the other, is subtracted from $ab$, the determinant of $L$ in (\ref{detL}) can be interpreted as the contribution from the destructive interference of the received fields at both sensors.  We now discuss the relationship between the two eigenvalues.

%%%%%%%%%%%%%%%%%%
\section{Connection to SEM}
\label{connectSEM}

Carl Baum first introduced SEM in the context of electromagnetic scatterers and antennas \cite{baum_toward_engineering_theory,baum88}. His approach to modeling the scattered field in terms of poles was originally motivated by the observation of typical transient responses of complex scatterers.  He conjectured that the responses were dominated by several damped sinusoids, and his main focus was scatterers on which the incident field creates surface current densities.  These densities can be analyzed in terms of the object's natural frequencies $\omega_\alpha$ \cite{baum88}, which typically occur at complex (non-physical) values.

In particular, each component of the scattering matrix $S$ is a meromorphic function of $\omega$, where the poles 
$\{\omega_{\alpha_k}\}_{k=1}^{\infty}$ satisfy  $\xi = 1 - q_1 q_2 \omega_{\alpha_k}^4 \tilde{G}^2(\vb{d},\omega_{\alpha_k}) =0$ and are put in order of increasing imaginary part.  Each element of the 
scattering matrix can be expanded in a Laurent series for $S(\omega)$ around a given pole $\omega_{\alpha_k}$  
\begin{equation}
S(\omega) = \sum\limits_{n=-m}^{\infty} s_n (\omega - \omega_{\alpha_k})^{n},
\end{equation} where $m$ is a positive integer, and $s_n$ is a matrix of the residues of the element of $S(\omega)$ at $\omega_{\alpha_k}$.  

%SEM considers the poles to be simple \cite{baum_toward_engineering_theory}, that is $(\omega-\omega_\alpha)S(\omega)$ is analytic around the neighborhood of $\omega_\alpha$.  
When the transmitted waveform $\tilde{\mathbf T}(\omega)$ is analytic, then $\tilde{\mathbf R}(\omega)$ is meromorphic.  
Moreover, both $\mathbf T(t)$ and $\mathbf R(t)$  must be causal ($\mathbf T(t) = \mathbf 0 = \mathbf R(t)$ for $t < 0$).
By the Paley-Wiener theorem \cite[p.~494]{lax}, both $\tilde{\mathbf T}(\omega)$ and $\tilde{\mathbf R}(\omega)$ are then analytic in the lower half-plane.  
Consequently the poles of the $\tilde{\mathbf{R}}(\omega)$ must lie in the upper half plane.  Using the Paley-Wiener Theorem allows one to avoid the convergence issues inherent in Baum's Laplace formulation.

To predict the time-domain received signal, we must compute the inverse Fourier transform \eqref{inversetransform} for the received signal, which may require modification of the integration contour.  
By an appropriate contour deformation and residue calculus,  
%In particular, we make an approximation of the inversion by moving the inversion line down to some $\Im(\omega_{\alpha_N})$, and we get
\begin{align}
\label{residue}
\vb{R}(t) \approx 2 \pi j \sum\limits_{i=1}^N & \underset{\omega=\omega_{\alpha_i} }{\mathrm{Res}} \left( \frac{\vb{\tilde{R}}(\omega) e^{j \omega t}}{2 \pi}  \right) \nonumber\\ & + O(e^{-|\mathrm{Im}(\omega_{\alpha_N})|t}),
\end{align} where the higher-order terms vanish as $t \to \infty$, and the notation $f(x) = O(g(x))$ means that  $|f(x)| \leq M |g(x)| $ for some positive real number $M$ as $x \to \infty$.  Equation (\ref{residue}) is the  SEM representation of the transient backscattered response at the receivers \cite{baum_toward_engineering_theory,ramm_SEM_foundation}.  

We note that the poles due to $\xi$ are a result of the interaction between the fields and scatterers.  These poles are aspect independent, and it is for this reason that they  have been proposed for use in target classification \cite{baum_target}.  Consequently, a number of methods, such as the Matrix Pencil Method and Prony's Method, have been developed for determining these poles \cite{sarkar_MPM_SEM_poles,sarkar,lobos,hurst_scattering_prony,hong}.  
Unfortunately, the problem of obtaining these poles from real data is ill-posed and sensitive to noise.  

Previous research \cite{dolph,marin} has shown, however, that there is a connection between the eigenvalues of the scattering matrix and the complex poles used for Baum's SEM.  Although the SEM poles are aspect-independent, unfortunately the scattering matrix itself and consequently the received signals are aspect-dependent.

%In particular, $\xi$ depends only by the physical parameters of the target -- reflectivity, scattering distance, and shape.  Though the interaction matrix (\ref{deltamatrix}) is aspect independent, the scattering matrix is aspect dependent, and the received signal is also dependent on aspect.  Consequently, the residues are also aspect dependent and is related to the scattering configuration.    The formulation of (\ref{residue}) is the class-1 form of the SEM representation \cite{baumPhillips}.  
%The aspect independent poles of the scatterers provide distinctive characterization of the target and have useful applications such as target classification \cite{baum_target}, and they can be determined by various methods  \cite{sarkar_MPM_SEM_poles,hongs,sarkar}.  

We show below that the TR matrix can be used to obtain certain information about the poles of $S$ in a stable manner.  First, we introduce the notation  $\hat{S}$: 
\begin{equation}	\label{hatS}
S = \frac{\omega^2}{\xi} P \Delta P^T = \frac{\omega^2}{\xi} \hat{S},
\end{equation} where $\hat{S} = P \Delta P^T$ and
\begin{equation}	\label{determinants}
\det(\hat{S}(\omega)) = \det(P(\omega)) \underbrace{\det(\Delta(\omega))}_{ q_1 q_2 \xi (\omega)} \det(P^T (\omega)).
\end{equation} 
The corresponding modified TR matrix $\hat{L}$ is
\begin{equation}	\label{hatL}
\hat{L} = \hat{S}^\dagger \hat{S}, \quad L = \left|\frac{\omega^2}{\xi}\right|^2 \hat{L}.
\end{equation}

From \eqref{singularity-xi}, at a natural frequency $\omega = \omega_\alpha$, we have $\xi(\omega_\alpha) = 0$; from  \eqref{determinants} we see that the determinant of $\hat{S}$ and therefore also of $\hat{L}$ must be zero, implying that at least one of the eigenvalues of $\hat{S}$ and $\hat{L}$ must be 0.  As $\omega \to \omega_\alpha$ and $\xi \to 0$, the eigenvalues of the interaction matrix $\Delta$ can be approximated  from (\ref{eigenvalues}) by 
\begin{eqnarray}
\lim\limits_{\omega \to \omega_\alpha}\lambda_1 (\omega) &\approx& q_1 + q_2, \\
\lim\limits_{\omega \to \omega_\alpha}\lambda_2 (\omega)&\approx& \frac{q_1 q_2}{q_1 + q_2} \xi, \label{lambda2}
\end{eqnarray}  
where the derivation of \eqref{lambda2} is in Appendix \ref{lambda2derivation}.  From (\ref{eigenvalueofS}) in Appendix \ref{appendMmatrix}, the eigenvalues of $S$ behave as
%\begin{eqnarray}
\begin{align}
\lim\limits_{\omega \to \omega_\alpha}\gamma_1 (\omega) &\approx \lim\limits_{\omega \to \omega_\alpha}\frac{\omega^2}{\xi} \lambda_1 m_{11} 
	\nonumber\\ &= \frac{\omega_\alpha^2}{\xi} (q_1 + q_2) m_{11}, \\
\lim\limits_{\omega \to \omega_\alpha} \gamma_2 (\omega) &\approx \lim\limits_{\omega \to \omega_\alpha} \frac{\omega^2}{\xi} \lambda_2 m_{22} \nonumber \\ &= \frac{\omega_\alpha^2 q_1 q_2 }{q_1 + q_2 }  m_{22}.
%\end{eqnarray} 
\end{align}
Clearly, $\gamma_1$ diverges as $\xi \to 0$, while $\gamma_2$ converges to 
$\omega_\alpha^2 q_1 q_2  m_{22} / (q_1 + q_2)$ .  
%Consequently the eigenvalues of the TR matrix behave as
%\begin{align}
%\lim_{\omega \to \omega_\alpha} \sigma_1 &= \frac{ \left| \omega^2 (q_1 + q_2) m_{11}   \right|^2}{|\xi|^2} \cr
%\lim_{\omega \to \omega_\alpha} \sigma_2 &= \left| \frac{\omega^2 q_1 q_2 }{u_1 + q_2 }  m_{22} \right|^2 
%\end{align}

We determine the eigenvalues of the TR matrix $L$ from $\{\gamma_1, \gamma_2 \}$ by (\ref{eigenvaluetrgamma}) in Appendix \ref{appendsvd}.  Consequently, the eigenvalues of $L$ behave as
\begin{align}
\lim\limits_{\omega \to \omega_\alpha}  \sigma_1 &= \left(\frac{\omega_\alpha^2}{\xi} (q_1 + q_2) m_{11} \frac{(\vb{v}_1^*,\vb{y}_1)}{(\vb{v}_1,\vb{y}_1)} \right)^2, \\
\lim\limits_{\omega \to \omega_\alpha} \sigma_2 &= \left(\frac{\omega_\alpha^2 q_1 q_2 m_{22} (\vb{v}_2^*,\vb{y}_2)}{(q_1 +u_2 )(\vb{v}_2,\vb{y}_2)} \right)^2,
\end{align} 
where $\vb{v}_i$ is an eigenvector of $L$ for $\sigma_i$ and $\vb{y}_i$ is an eigenvector of $S$ for $\gamma_i$. 
Clearly, the larger eigenvalue of $L$ diverges to positive infinity in the limit $\omega \to \omega_\alpha$, while smaller eigenvalue $\sigma_2$ converges to a finite number.   
Because in general the natural frequencies $\omega_{\alpha_k}$ are complex-valued, the larger eigenvalue does not actually diverge for real frequencies, and instead one sees resonance peaks.  

%That eigenvalue $\sigma$ which corresponds to the pole  $\omega_\alpha$ diverges because  $\xi \to 0$ as $\omega \to \omega_\alpha$.  For a given complex frequency $\omega =  2 \pi \nu -  j \rho$, where $\nu$ is the frequency and $\rho$ is the damping term, we write this complex frequency in the polar form 
%\begin{equation}
%\label{complexfrequency}
%\omega = \omega_0 \exp (j \Phi ).
%\end{equation}  The amplitude $\omega_0 = \sqrt{4 \pi^2 \nu^2 + \rho^2}$, and $\Phi$ is the phase of $\omega$.  We re-write $\xi$ in (\ref{xidef}) as 
%\begin{equation}
%\label{xicomplexfreq}
%\xi = 1 -\frac{\mu_1 \mu_2 \omega_0^4 e^{-2 \omega_0 d  \sin \Phi/c} e^{-j(2 \omega_0 d \cos \Phi/c -4 \Phi)}}{16 \pi^2 d^2}.
%\end{equation}  For a given damping term $\rho$, $\xi$ can be minimized by $\omega_0$ in (\ref{xicomplexfreq}).  

When the damping is small ($\mathrm{Im} (\omega_\alpha)<<1$), the real resonant frequency is approximately equal to the complex natural frequency, and we will expect a significant response at this frequency.  The difference between a real resonant frequency $\omega_\alpha$ and its associated complex natural frequency depends on the phase of $\omega_\alpha$.  

Thus the TR algorithm provides information about the rotation of the natural frequencies $\omega_{\alpha_k}$ onto the real axis. 
%Moreover, since the TR algorithm is stable and gives rise to increased  energy received from target scattering, use of the TR algorithm is likely to provide this information in a reliable manner.  
Moreover, we show in the next section that this information can be obtained in a stable manner.  
Therefore, the TR algorithm provides a stable way of obtaining information about the SEM poles.  

%In the following section, we compare the stability of the eigenvalues of the TR and the Scattering matrices.

%%%%%%%%%%%%%%%%%%%%%%%%%
\section{Stability of  Eigenvalues}
\label{eigenvaluesstability}
%%%%%%%%%%%%%%%%%
In this section, we  compare and contrast the eigenvalues of $S$ and $L$ and show how these eigenvalues are respectively perturbed by noise.

%%%%%%%%%%%%%%%
\subsection{Stability of finding eigenvalues from measurements of $S$}
Thermal noise gives rise to perturbations in the elements of the scattering matrix.  
We would like to examine how a small change $\delta S$ in the scattering matrix is translated to the changes $\delta \Gamma$ in the eigenvalues of $S$. 
We re-write (\ref{eigendecompS}) as 
\begin{equation}
\label{invertedscattering}
Y^{-1} S Y = \Gamma.
\end{equation} 
Inserting $\Gamma+\delta \Gamma$ and $S+\delta S$ into (\ref{invertedscattering}) and rearranging, we have 
\begin{equation}
\delta \Gamma = Y^{-1} \delta S Y. 
\end{equation} 
Taking the operator norm $\| \cdot \|$ and applying a standard matrix product inequality yields 
\begin{equation}
\| \delta \Gamma \| = \|Y^{-1} \delta S Y \| \leq \|Y^{-1} \| \|\delta S \| \|Y\|.
\end{equation}
%Since the matrices depend on $\omega$, define the induced norm at the frequency that gives the highest upper bound.  
The error is then bounded by 
\begin{equation}
\max_\omega \|\delta \Gamma \| \leq \|Y^{-1} \| \|Y\| \|\delta S\| = \max_\omega \: \kappa(Y) \| \delta S \|,
\end{equation} 
where $\kappa$ is the condition number.  The errors in the eigenvalues are thus bounded by the errors in measurement introduced in $S$ and the conditioning of the eigenvectors of $S$ (columns of $Y$).  Since $Y$ depends on the field propagation paths and scattering, the conditioning will be a function of the configuration of the transmitters, receivers, scatterers, and the frequency.  In general, finding the eigenvalues of $S$ from measurements of $S$ is an unstable process.  

%%%%%%%%%%%%%%%%%
\subsection{Stability of finding eigenvalues from TR measurements}
For the TR operator, the errors in the eigenvalues are less subject to the configuration of the system.  From (\ref{TRdecomp}), 
\begin{equation}
\max_\omega \|\delta \Sigma \| \leq \|V\| \|V^\dagger\| \| \delta L\|.
\end{equation}Since $V$ is unitary, the norms are both $1$, and 
\begin{equation}
\label{errorbound}
\max_\omega \|\delta \Sigma \| \leq \max_\omega \|\delta L\|.
\end{equation}
The error in the eigenvalues of $L$ is bounded by the error introduced in the measurement and does not rely on the conditioning of $Y$.  The error bound in (\ref{errorbound})  suggests  that finding the eigenvalues of  the TR operator is a more well-conditioned problem than finding the eigenvalues of the scattering operator and may provide a stable way of calculating the eigenvalues of the scattering operator. 
%

%%%%%%%%%%%%%%%%%%%%%%%%%
\section{Numerical Results}
\label{numericalresults}

In this section, we provide simulations of the TR algorithm for two transmitter/receiver pairs and two isotropic scatterers in free space.  The two transmit/receive pairs are placed at coordinates  (-500 m, 700 m, 10 m) and (-500 m, 1000 m, 10 m).  The scatterers are separated by 500 m, with the strong one $(q_1)$ located at (150 m, 850 m, 10 m) and the weaker one $(q_2)$ located at (-107.52 m, 1278.58 m, 10 m).  In addition, the transmitter/receiver pairs and the scatterers are placed in a plane with $q_1 = 0.75 \: \mathrm{Hz}^{-2} \mathrm{m}^{-2}$ and $q_2=0.4 \: \mathrm{Hz}^{-2} \mathrm{m}^{-2}$.  The initial waveform $T^0(t)$ for each transmitter is the same linear frequency modulated (LFM) signal with the following characteristics:
%\begin{figure}[!h]
%\centering
%\includegraphics[width=3.5in]{./figures/picture1.png}
%\caption{Two transmit/receive pairs are placed at coordinates  (-500 m,700 m,10 m) and (-500 m,1000 m,10 m).  The cylinder represents the strong scatterer $q_1$ located at (150m,850m,10m).  The sphere represents the weaker scatterer $q_2$ located at (-107.52m,1278.58m,10m).  The scatterers are 500m apart.}  
%\label{radar:configuration}
%\end{figure}
%=75 \times 10^8 \mathrm{Hz}^{-2} \mathrm{m}^{-2}
%= 4 \times 10^8 \mathrm{Hz}^{-2} \mathrm{m}^{-2}

\begin{enumerate}
\item $10 \: \mu \mathrm{s}$ pulse duration;
\item $10^{12} \: \mathrm{s}^{-2}$ chirp rate;
\item $100 \: \mathrm{MHz}$ carrier frequency.
\end{enumerate}

\begin{figure}[!h]
\centering
\includegraphics[width=3.5in]{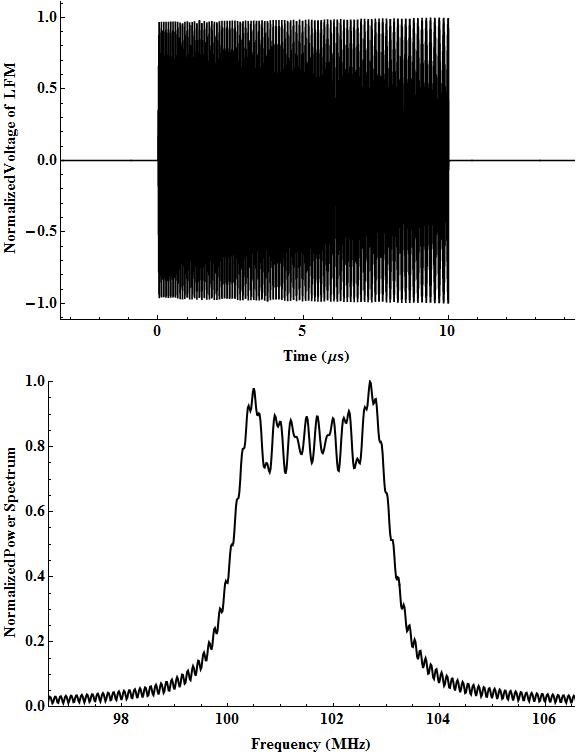}
\caption{Original LFM Signal $T^0$: time-domain signal (top); square magnitude of power spectrum (bottom) with 4 MHz bandwidth centered at 100 MHz.}  
\label{chirpsignal}
\end{figure}

For the simulations, we limit ourselves to the even iterations of the transmitted signal. We normalize the signal vector with the euclidean norm for each iteration.  We observe the transformation of the original chirp to the final waveform in the time domain and the power spectrum of the signal.
\begin{figure}[!h]
\centering
\includegraphics[width=2.4in]{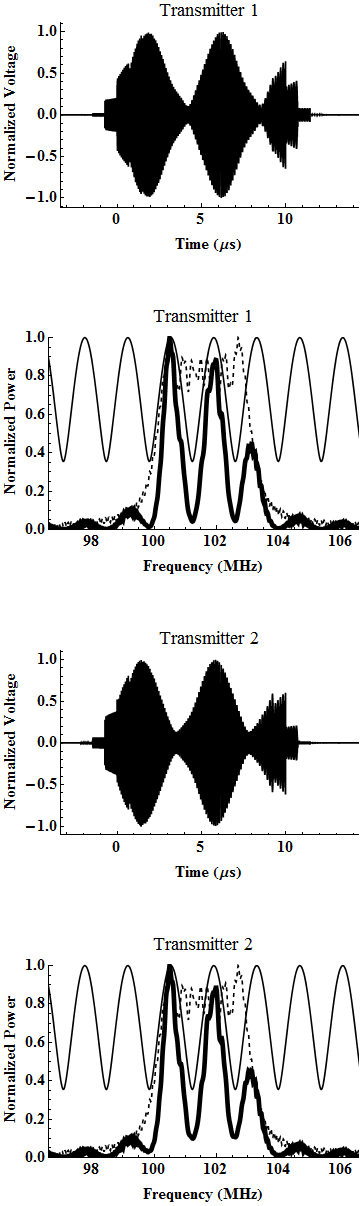}
\caption{Top two rows show the normalized transmitted waveform and the power spectrum of sensor 1 after two iterations of the TR algorithm.  Bottom two rows show the normalized transmitted waveform and power spectrum of sensor 2 after two iterations.  The original LFM (dotted line), the normalized power spectrum of the signal (thick line), and the normalized calculated eigenvalues (thin line) are shown.  Scatterers are 500m apart. 
}  
\label{configuration_largestrength}
\end{figure}
Figure \ref{configuration_largestrength} shows the final waveform from $T^0$ after 2 iterations of the TR process, the normalized magnitude of the spectra for TR-generated signals, and the normalized calculated eigenvalues of $L$ for a scatterer separation of 500 m.  As the figure indicates, the time-domain waveform significantly changes and broadens from the original LFM signal for both transmitters.  The power spectrum of each new waveform has relative maximum values at the resonant frequencies.  Observe that for a given bandwidth of the original LFM signal, the energy of the TR-generated signals after 2 iterations (thick line) concentrates near the resonant frequencies.  Significantly, the maxima of this power spectrum aligns with the maxima of the calculated eigenvalues of the TR operator (thin line).  The peaks of the power spectrum occur within the bandwidth of the power spectrum of $T^0$ (dashed curve).   

\begin{figure}[!h]
\centering
\includegraphics[width=2.4in]{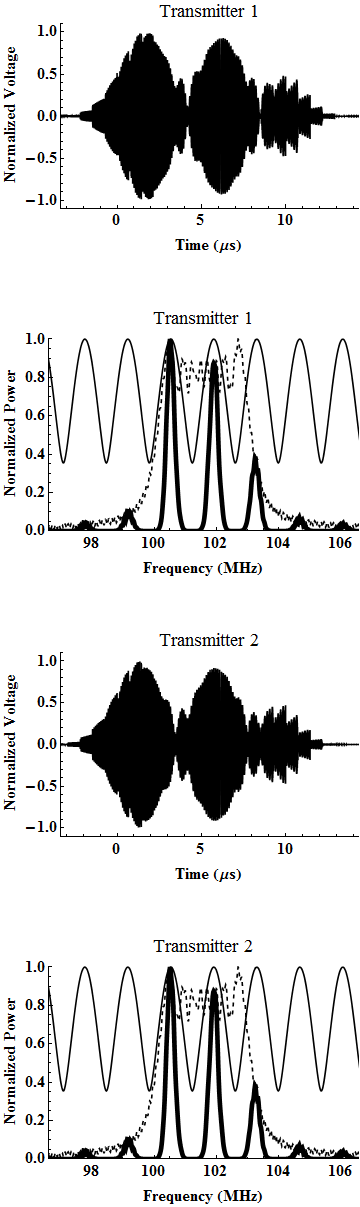}
\caption{Top two rows show the normalized transmitted waveform and the power spectrum of sensor 1 after ten iterations of the TR algorithm.  Bottom two rows show the normalized transmitted waveform and power spectrum of sensor 2 after ten iterations.  The original LFM (dotted line), the normalized power spectrum of the signal (thick line), and the normalized calculated eigenvalues (thin line) are shown.  Scatterers are 500m apart. }  
\label{configuration_largestrength2}
\end{figure}
At higher iterations of the TR process, the field intensities increase at the resonant frequencies  and decrease at other frequencies..  In particular, Fig. \ref{configuration_largestrength2} plots the power spectrum and the time-domain signal at 10 iterations.  The frequency peaks are sharp, and one can observe the key resonating frequency where the intensity is at its highest. The time-domain waveform has  broadened from the $10 \: \mu s$ pulse in Fig. \ref{chirpsignal} as the new waveform is determined by the resonant frequencies.  The plot of the power of the original LFM signal is now narrowed and centered around the resonant frequencies.  The largest intensity is found within the bandwidth of the original signal.

\begin{figure}[!h]
\centering
\includegraphics[width=3in]{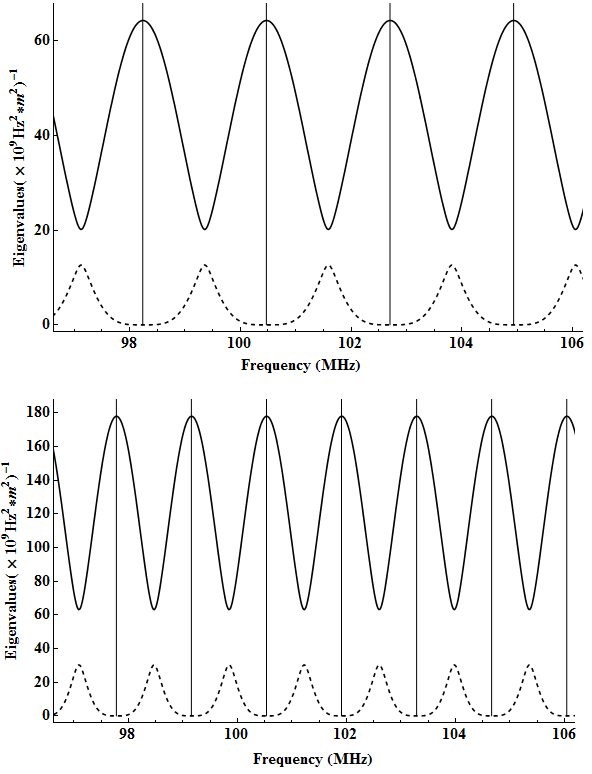}
\caption{First and second eigenvalues of the TR matrix $L$ are shown at two separate distances of the point scatterers, 300 m (top) and 500 m (bottom) respectively with the transmitter and stronger scatterer in the same position.  The two figures shown are the theoretical eigenvalues -- larger eigenvalue (solid), smaller eigenvalue (dashed), and the vertical line indicates the Bragg frequencies.}
\label{configuration_largestrength3}
\end{figure}

Figure \ref{configuration_largestrength3} shows the large and small eigenvalues.  As discussed in Section \ref{connectSEM}, the larger eigenvalues attain relative maximum values at resonance.  The vertical lines indicate the frequency at which Bragg's condition is met.  
Maximum constructive interference occurs at these frequencies and matches the eigenvalue peaks which are $1.37$ MHz apart.  This frequency separation corresponds to the projected distance between the scatterers, where the projection is onto the axis normal to the axis vector from the strong scatterers to the midpoint of the sensors ($d_{\mathrm proj}$ in Fig. \ref{bragg_geometry}).  In this case, the actual distance between the scatterers is $500$ m ($=$$d$), and the projected distance is $484$ m.
For both scatterer separations in Fig. \ref{configuration_largestrength3}, the larger eigenvalue has a maximum at the resonance frequencies.  The locations of the resonances (vertical lines) are exact because the formula for Bragg's Law  (a geometric approximation) is not used.  However, when Bragg's Law  is used, the vertical lines will be slightly displaced from the relative maxima due to the geometrical approximation introduced in deriving the Bragg's formula in Appendix \ref{append:bragg}.  This displacement error can be explained by modifying the analysis treated in \cite{mokole}.

\begin{figure}[!h]
\centering
\includegraphics[width=2.6in]{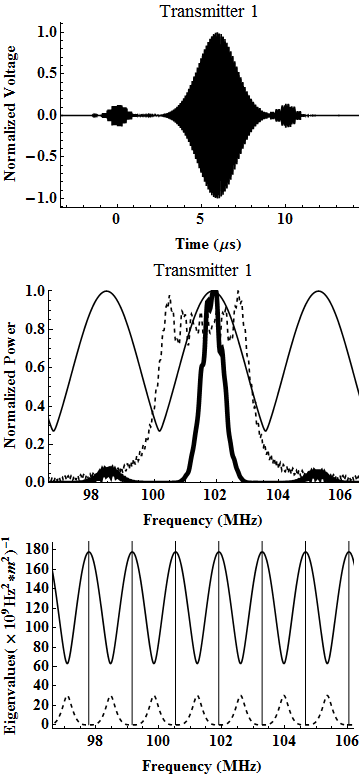}
\caption{The scatterers rotated $81^\circ$ in azimuth from the initial configuration.  $d$ remains the same.  The eigenvalue peaks are $3.39$ MHz apart.  TR process was taken out to 10 iterations.  Top row is the time-domain signal; middle row is the power spectrum, bottom row is the first (solid line) and second eigenvalues (dashed line) of the TR matrix.}  
\label{degreeshift}
\end{figure}
%{\bf **** again, change the lines as in above figures and reference each line*****}
In Figure \ref{degreeshift}, the weaker scatterer is rotated $81^{\circ}$ in azimuth about the stronger scatterer from the original configuration.  The projected distance is $197$ m.  The scatterer separation distance of $500$ m remains the same, but the separation between frequency peaks changes to 3.39 MHz.  As the projected distance decreases, the eigenvalue peaks will be further apart.  A signal with a broader bandwidth will be required to observe multiple resonant peaks.  Thus the aspect angle, which determines the propagation paths of the fields, has a strong influence on the locations of the eigenvalue peaks.

%%%%%%%%%%%%%%%%%%%%%%%%%
\section{Conclusions}
\label{conclusion}

We have shown that in a multiple-scattering environment,  the TR matrix has eigenvalues that vary with frequency in ways that can be predicted by the interaction of the scattered waves.   In contrast to the case of non-interacting scatterers \cite{prada,prada_eigenmodes} where different eigenvalues correspond to different scatterers, we have shown that when multiple scattering is present,  each eigenvalue involves the field interactions from all the propagation paths and all the scatterers.  The largest eigenvalue can be thought of as corresponding to the field intensities at the receivers from all the scattered waves.  The frequencies at which the largest eigenvalue has local maxima are the resonance frequencies, where maximum constructive interference occurs, and the frequency difference between the peaks depends on the wavelength and geometry of the scatterers. 

Our work shows that the TR process provides a stable way of finding the resonances of a scattering system.   Moreover, the TR process can be applied to an arbitrary waveform of arbitrary bandwidth.  Our work confirms the findings of  \cite{cheney_electromagnetics,cheneyoptimal}, which showed, for an idealized half-space scattering geometry, the TR algorithm automatically converges to the space-time waveform that maximizes the energy scattered back to the receivers.   Our work shows that the same result holds for a more realistic scattering geometry.  Thus the best waveform for target detection is a single harmonic waveform at one of the resonance frequencies, and the TR process provides a stable method to find these resonance frequencies.

We have shown, moreover, that these scattering system resonances are closely related to the poles of the SEM.  In particular, as the poles of the SEM approach the real axis, the SEM poles become exactly the resonances found by the TR process.  The precise nature of the information about the SEM poles that is provided by the TR process is a question we leave for future work.

%% file: sections/appendix.tex
\appendices
%% you can choose not to have a title for an appendix
%% if you want by leaving the argument blank

%
\section{Derivation of the eigenvalues of the Scattering Matrix $S$}
\label{appendMmatrix}
The characteristic polynomial of (\ref{scatteringmatrixeigen}) using (\ref{Mmatrix}) can be expressed as 
\begin{equation}
\label{detofchar}
\det \left( \frac{\omega^2}{\xi} \Lambda M - \gamma I \right) = 0,
\end{equation} 
\begin{equation}
\label{lambdaandM}
\Lambda = \left[\begin{array}{cc}
\lambda_1 & 0 \\ 0 & \lambda_2
\end{array}\right], \quad M = \left[\begin{array}{cc}
m_{11} & m_{12} \\ m_{21} & m_{22}
\end{array}\right],
\end{equation}  which after substitution of (\ref{lambdaandM}) becomes
\begin{equation}
\label{moredet}
\det \left(\left[\begin{array}{cc}
\frac{\omega^2}{\xi}\lambda_1 m_{11} - \gamma & \frac{\omega^2}{\xi} \lambda_1 m_{12} \\ \frac{\omega^2}{\xi} \lambda_2 m_{21} & \frac{\omega^2}{\xi} \lambda_2 m_{22} - \gamma 
\end{array}\right]\right) = 0.
\end{equation} Since the matrix is $2 \times 2$, the solution to (\ref{moredet}) can be expressed in terms of the trace and determinant of $\Lambda M$:  
\begin{equation}
\gamma =\frac{\omega^2}{\xi}  \frac{\mathrm{tr} (\Lambda M) \pm \sqrt{\mathrm{tr} (\Lambda M)^2 - 4 \det(\Lambda M)} }{2},
\end{equation} where 
\begin{eqnarray*}
\mathrm{tr}(\Lambda M) &=& \lambda_1 m_{11} + \lambda_2 m_{22} \\
\det(\Lambda M) &=& \lambda_1 \lambda_2 m_{11} m_{22} - \lambda_1 \lambda_2 m_{12} m_{21}.
\end{eqnarray*}  From (\ref{Mmatrix}), $m_{ip}$  can be written in index form as 
\begin{equation}
m_{ip} = \sum\limits_{n=1}^{2} \sum\limits_{k=1}^{2} \sum\limits_{l=1}^{2} X_{in}^{-1} P_{kn}P_{kl},X_{lp},
\end{equation} and
\begin{align}
\label{eigenvalueofS}
 \gamma_{1,2} &= \frac{\omega^2}{\xi} \times \left( \frac{\lambda_1 m_{11} +\lambda_2 m_{22}}{2} \right. \nonumber \\ & \left. {\pm \sqrt{(\lambda_1 m_{11}-\lambda_2 m_{22})^2 + 4 \lambda_1 \lambda_2 m_{21}m_{12}}\over 2} \right).
\end{align}

%%%%%%%%%%%%%%%%%%%%%%%%%
\section{Relationship between Singular Values of the Scattering Operator $S$ and the Eigenvalues of the TR Operator $L$} \label{appendsvd}

%Any matrix $S$ can be written in terms of its singular value decomposition.  
%If 
%\begin{equation}
%S \mathbf{v} = \zeta \mathbf{u} \quad \mathrm{and} \quad S^\dagger \mathbf{u} =\zeta \mathbf{v},
%\end{equation} 
%where $\zeta$  is a real scalar, and $\mathbf{u}$ and $\mathbf{v}$ are non-zero complex vectors, 
%then $\zeta$ is called a singular value of $S$ and $\mathbf{u}$ and $\mathbf{v}$ are called the  left and right singular vectors, respectively.  The set of such vectors are orthogonal and form a basis of their respective linear space such that the left singular vectors form a unitary matrix $U$, and the right singular vectors form a unitary matrix $V$.  If we denote by $\Sigma$ the diagonal matrix consisting of the singular values $\zeta_i$, then 

Any matrix $S$ has a singular value decomposition
\begin{equation}	\label{SVD}
S= U \Sigma V^\dagger,
\end{equation} 
where $U$ and $V$ are unitary matrices.   If we multiply \eqref{SVD} by the adjoint $S^\dag$, then we have
\begin{eqnarray}
L = S^\dagger S &=& V \Sigma^\dagger U^\dagger U \Sigma V^\dagger\\ &=& V \Sigma^2 V^\dagger,
\end{eqnarray}  and then post multiplying $V$ on both sides implies
\begin{equation}
S^\dagger S V = V \Sigma^2.
\end{equation}  Therefore, the right singular vectors of $S$ are the eigenvectors of $L=S^\dagger S$, and the eigenvalues $\{\sigma_1,\sigma_2\}$ of $L$ are the squares of the corresponding singular values $\{\zeta_1,\zeta_2\}$ of $S$.  
  
For a square matrix with distinct singular values, the left and right singular vectors are unique up to a complex sign ({\it i.e.}, complex scalar factors of absolute value unity) \cite[p.~29]{trefethen}.  Since $S$ is square and symmetric but not self-adjoint, the left singular vectors are the conjugates of the right singular vectors:
\begin{equation}
\label{newS}
V^* \Sigma U^T = S^T = S =U \Sigma V^\dagger.
\end{equation} Therefore $U=V^*$, which implies that for any right singular vector $\vb{v}_l$, there is a left singular vector $\vb{u}_l$ such that  
\begin{equation}		\label{Sv}
S \mathbf{v}_l=\zeta_l \mathbf{u}_l = \zeta_l \mathbf{v}_l^*.
\end{equation}

The inner product of $L \vb{v}_l$ with an eigenvector $\mathbf y_n$ of $S$ corresponding to the eigenvalue $\gamma_n$ is
\begin{align}		\label{evectorRelation}
 \sigma_l (\mathbf{v}_l, \mathbf{y}_n) &= (L \mathbf{v}_l,\mathbf{y}_n) = (S^\dagger S \mathbf{v}_l,\mathbf{y}_n) 
 	= (S \mathbf{v}_l, S \mathbf{y}_n)
\end{align} 
where $\zeta_l^2 = \sigma_l$.  
In \eqref{evectorRelation}, we use expression (\ref{Sv}) for $S \mathbf{v}_l$ and the relation $S \mathbf y_n = \gamma_n \mathbf y_n$ to obtain
\begin{equation}
\sigma_l (\mathbf{v}_l,\mathbf{y}_n) = \zeta_l \gamma_n (\mathbf{v}_l^* , \mathbf{y}_n) .
\end{equation}  Since $\sqrt{\sigma_l} = \zeta_l$,
\begin{equation}
\label{eigenvaluetrgamma}
\sqrt{\sigma_l}  = \gamma_n \frac{(\mathbf{v}_l^*,\mathbf{y}_n)}{(\mathbf{v}_l,\mathbf{y}_n)}.
\end{equation}

%%%%%%%%%%%%%%%%%%%%%%%%%%%
\section{Derivation of the TR matrix $L$}
\label{appendvalues}
We recall the definitions \eqref{hatS}, \eqref{propagatorback}, and \eqref{deltamatrix} %(85), (29), and (31)  in July 9 version
of $\hat S$, the propagation matrix $P$,  
%\begin{equation}
%P = \left[\begin{array}{cc}
%\tilde{G}(\mathbf{y}_1-\mathbf{X}_1, \omega) & \tilde{G}(\mathbf{y}_1- \mathbf{X}_2,\omega) \\
%\tilde{G}(\mathbf{y}_2- \mathbf{X}_1,\omega) & \tilde{G}(\mathbf{y}_2-\mathbf{X}_2,\omega)
%\end{array}\right]
%\end{equation}
and the interaction matrix, 
%\begin{equation}
%\left[ \begin{array}{cc}
%  u_1 & u_1 u_2 \omega^2 \tilde{G}(d,\omega)  \\
%  u_1 u_2 \omega^2 \tilde{G}(d,\omega) & u_2 
%  \end{array}\right]
%\end{equation} determines the scattering matrix
together with the relation $\hat{S} = P \Delta P^T $ between them.  
%\end{equation} and 
Then in index form, 
\begin{equation}
\label{Ssum}
\hat{S}_{ij} = \sum_k \sum_l P_{ik} \Delta_{kl} P_{jl}.
\end{equation}  
The corresponding modified TR matrix, which is given by \eqref{hatL}, is
\begin{equation}
 \hat{L}_{mj} = \sum_i \hat{S}^*_{im} \hat{S}_{ij}.
\end{equation}  The trace of $\hat{L}$ is
\begin{equation}
\mathrm{tr} (\hat{L}) = \hat{L}_{11} + \hat{L}_{22},
\end{equation} where
\begin{eqnarray}
\hat{L}_{11} &=& \sum_i \hat{S}^*_{i1} \hat{S}_{i1},\\
\hat{L}_{22} &=& \sum_i \hat{S}^*_{i2} \hat{S}_{i2}.
\end{eqnarray}  

The elements of $S$ using (\ref{Ssum}) are 
\begin{align}	
\hat{S}_{11} = P_{11} \Delta_{11} P_{11} &+ P_{11} \Delta_{12} P_{12} +P_{12} \Delta_{21} P_{11} \nonumber \\ &+ P_{12} \Delta_{22} P_{12}, \\
\hat{S}_{22} = P_{21} \Delta_{11} P_{21} &+ P_{21} \Delta_{12} P_{22} + P_{22} \Delta_{21} P_{21} \nonumber \\ &+ P_{22} \Delta_{22} P_{22}, \\
\hat{S}_{12} = P_{11} \Delta_{11} P_{21} &+ P_{11} \Delta_{12} P_{22} + P_{12} \Delta_{21} P_{21} \nonumber \\ &+ P_{12} \Delta_{22} P_{22}, \\
\hat{S}_{21} = P_{21} \Delta_{11} P_{11} &+ P_{21} \Delta_{12} P_{12} + P_{22} \Delta_{21} P_{11} \nonumber \\ &+ P_{22} \Delta_{22} P_{12},
\end{align} which gives \eqref{Sabc}, \eqref{a}, \eqref{b}, and \eqref{c}.  
%Since $S=S^T$, $S_{12} =S_{21}$, and we define $a = S_{11}$, $b=S_{22}$, and $c=S_{12}=S_{21}$.  We can interpret $a$ and $b$ as the response of the system in the $i$ and $j$ sensor, and $c$ as the cross-sensor response from $i$ to $j$.
%\begin{align}
%\label{sa}
%S_{11} =  a = & \left(u_1 \tilde{G}^2 (\vb{y}_1-\vb{x}_1) \right. \nonumber \\  + & 2 u_1 u_2 \omega^2 \tilde{G}(d) \tilde{G}(\vb{y}_1 - \vb{x}_1) \tilde{G}(\vb{y}_1 - \vb{x}_2) \nonumber \\ & \left. +u_2 \tilde{G}^2 (\vb{y_1}-\vb{x}_2) \right) \frac{\omega^2}{\xi},
%\end{align}  
%\begin{align}
%\label{sb}
%S_{22} = b = & \left(u_1 \tilde{G}^2(\vb{y}_2 - \vb{x}_1) \right. \nonumber \\  + & 2 u_1 u_2 \omega^2 \tilde{G}(d) \tilde{G}(\vb{y}_2 - \vb{x}_2) \tilde{G}(\vb{y}_2 - \vb{x}_2) \nonumber \\
%& \left. + u_2 \tilde{G}^2(\vb{y}_2 - \vb{x}_2) \right) \frac{\omega^2}{\xi},
%\end{align} and
%\begin{align}
%\label{sc}
%S_{12} =&  S_{21} = c = \left(  u_1 \tilde{G} (\vb{y}_2 - \vb{x}_1)\tilde{G}(\vb{y}_1 - \vb{x}_1) \right. \nonumber\\
%+ & u_1 u_2 \omega^2 \tilde{G}(d) \tilde{G}(\vb{y}_1 -\vb{x}_1) \tilde{G}(\vb{y}_2 - \vb{x}_2) \nonumber \\ & + u_1 u_2 \omega^2 \tilde{G}(d) \tilde{G}(\vb{y}_1 - \vb{x}_2) \tilde{G} (\vb{y}_2 - \vb{x}_1) \nonumber \\ & \left. + u_2 \tilde{G}(\vb{y}_1 - \vb{x}_2) \tilde{G}(\vb{y}_2 - \vb{x}_2) \right) \frac{\omega^2}{\xi}.
%\end{align}
The  elements of the TR matrix are then
\begin{eqnarray}
L_{11} &=& |S_{11}|^2 + |S_{21}|^2 = |a|^2 + |c|^2\\
L_{22} &=& |S_{22}|^2 +|S_{12}|^2 = |b|^2 + |c|^2\\
L_{12} &=& S_{11}^* S_{12} + S_{21}^* S_{22} =a^* c + c^* b\\
L_{21} &=& S_{12}^* S_{11} + S_{22}^* S_{21} = c^*a + b^* c.
\end{eqnarray}  

\section{Derivation of Bragg's Formula}
\label{append:bragg}
From Figure \ref{scatteredfieldgeometry}, $\triangle ABC$ and the law of sines yield
\begin{eqnarray}
\frac{\sin \zeta}{l_1} &=& \frac{\sin \theta}{d}, \\
\label{acb}
\frac{\sin \theta}{d} &=& \frac{\sin \angle ACB}{l_3}, 
\end{eqnarray} and 
$\angle ACB = \pi -\theta - \zeta$.  Substituting this angle into \eqref{acb}, we obtain
\begin{eqnarray}
\frac{\sin \theta}{d} &=& \frac{\sin (\theta + \zeta)}{l_3}.
\end{eqnarray} For $\triangle DBC$,

\begin{eqnarray}
\frac{\sin \phi}{d} &=& \frac{\sin \psi}{l_4}, \\
\label{dcb}
\frac{\sin \phi}{d} &=& \frac{\sin \angle DCB}{l_2},
\end{eqnarray} and
$\angle DCB = \pi - \phi - \psi$.  Substituting this angle into \eqref{dcb}, we obtain
\begin{eqnarray}
\frac{\sin \phi}{d} &=& \frac{\sin (\phi + \psi)}{l_2}.
\end{eqnarray}  These equations lead to \eqref{l1}, \eqref{l2}, \eqref{l3}, and \eqref{l4}.  For $\triangle ABC$, the law of cosines yields
\begin{equation}
\frac{l_1^2 + l_3^2 -d^2}{2 l_1 l_3} = \cos \theta.
\end{equation}  As $l_1 \to \infty, l_3 \to \infty$, $\cos \theta \to 1$, and $\theta \to 0$, the second term of \eqref{newrelation} is
\begin{equation}
\lim\limits_{\theta \to 0} \frac{\sin(\zeta + \theta) - \sin \zeta}{\sin \theta}  = \cos \zeta.
\end{equation} 
For the first term of \eqref{newrelation}, we look at $\triangle DBC$:  from the law of cosines we have
\begin{equation}
\frac{l_2^2 +l_4^2 - d^2}{2 l_2 l_4}  = \cos \phi.
\end{equation}  As $l_2 \to \infty, l_4 \to \infty$, $\cos \phi \to 1$, and $\phi \to 0$, the first term of \eqref{newrelation} has the limit
\begin{equation}
\lim\limits_{\phi \to 0} \frac{\sin \psi - \sin (\phi + \psi)}{\sin \phi} = - \cos \psi.
\end{equation}  When the sensors are far from the scatterers, \eqref{newrelation} can be approximated as 
\begin{equation}
\label{approxrelationship}
d (- \cos \psi + \cos \zeta) = q \lambda_{\text{wave}} .
\end{equation} 
Moreover, since the sum of the interior angles of the triangle $\triangle DBC$ must be $\pi$, and the sum of the unmarked angle of $\triangle DBC$ plus $\eta$ plus $\beta$ must be $\pi/2$, and recalling that $\phi \approx 0$, we have 
%and the angles in $\triangle DBC$ and $\triangle ABC$ have the following relationship
\begin{eqnarray}
\psi &=& \frac{\pi}{2} + \eta + \beta. 
\end{eqnarray}
Similarly, we have 
\begin{eqnarray}
\zeta &=& \frac{\pi}{2} + \alpha + \eta.
\end{eqnarray}  In the far field, trigonometric identities can be used to rewrite (\ref{approxrelationship}) 
%in terms of the angles between the normal shown in Figure \ref{scatteredfieldgeometry} and the legs connecting from the reference scatterer to the sensors, yielding 
as (\ref{finalrelation}).

%
%%%%%%%%%%%%%%%%%%
\section{Derivation of \eqref{lambda2}} 	\label{lambda2derivation}
In \eqref{eigenvalues} we use \eqref{xidef} to replace $q_1 q_2 \omega^4 \tilde{G}^2(d,\omega)$:
\begin{align}
\lambda_2 &= {1 \over 2} \left[ q_1 + q_2 - \sqrt{ (q_1 - q_2)^2 + 4 q_1 q_2 ( 1 - \xi) }  \right] \cr
&= {1 \over 2} \left[  q_1 + q_2 -  \sqrt{ (q_1 + q_2)^2 - 4 q_1 q_2 \xi} \right] \cr
&=  {1 \over 2} \left[  q_1 + q_2 -   (q_1 + q_2) \right. \cr  & \hspace{2.5cm} \times \left. \left( 1 - {1 \over 2} {4 q_1 q_2 \xi \over  (q_1 + q_2)^2} + \cdots\right) \right] \cr
 &= {q_1 q_2 \xi \over  q_1 + q_2} + O \left(\frac{q_1^2 q_2^2 \xi^2}{(q_1 +q_2)^3}\right)
\end{align}

%\section{Extension to $N$ scatterers}
%%Appendix two text goes here.
%%
%
%
%For a more complex model, one can extend (\ref{deltamatrix}) to $N$ scatterers by setting 
%\begin{align}
%\label{bigdelta}
% & \Delta  = {\omega^2 \over \xi} \times \nonumber \\
% & \left[ \begin{array}{cccc}u_1 & u_1 u_2 \omega^2 \tilde{G}(L_{12})  & \ldots & u_1\mu_N\omega^2\tilde{G}(L_{1N})\\
%u_2 u_1 \omega^2 \tilde{G}(L_{12}) & u_2 &   \ldots & u_2 \mu_N\omega^2\tilde{G}(L_{2N}) \\
%\vdots & \vdots & \ddots & \vdots\\
%\mu_N u_1 \omega^2 \tilde{G}(L_{1N}) & \mu_N u_2 \omega^2 \tilde{G}(L_{2N}) & \ldots & \mu_N
%\end{array}\right]   
%\end{align} where $L_{ij}$ is the length between the scatterer $\mu_i$ and $\mu_j$, and $\xi$ is the determinant of  
%\begin{align}
%&\xi = \nonumber \\ &\left|\begin{IEEEeqnarraybox*}[][c]{,c/c/c/c,} 1 & -\omega^2u_2 \tilde{G}(L_{12}) & \ldots & -\omega^2 \mu_N \tilde{G}(L_{1N}) \\
%-\omega^2 u_1 \tilde{G}(L_{12}) & 1 & \ldots& -\omega^2 \mu_N \tilde{G}(L_{2N})\\
%\vdots & \vdots & \ddots & \vdots \\
%-\omega^2u_1\tilde{G}(L_{1N} & -\omega^2u_2\tilde{G}(L_{2M}) & \ldots & -\omega^2\mu_N\tilde{G}(L_{2N}) \end{IEEEeqnarraybox*}\right|.
%\end{align}  This accounts for the coupling between the point scatterers.  The propagator matrix $P$ will be a $M$ by $N$ matrix describing $M$ transmit receive pairs and $N$ scatterers.

%\section{yet another section}
%Appendix three text goes here.

%% file: sections/acknowledgment.tex
\ifCLASSOPTIONcompsoc
  \section*{Acknowledgments}
\else
  \section*{Acknowledgment}
\fi

The authors thank Dr. Sun Hong and Dr. Tim Andreadis from Tactical Electronics Warfare Division for their comments, feedback and support of this work.  The authors also like to thank Dr. Shannon Blunt from the University of Kansas for his inputs and relevant references.  The work of the first author was performed under the auspices of the Naval Research Laboratory base programs (approved for public release, distribution unlimited).  
M.C. is grateful to the Air Force Office of Scientific Research for support of this work under agreement FA9550-14-1-0185.\footnote{Consequently the U.S. Government is authorized to reproduce and distribute reprints for Governmental purposes notwithstanding any copyright notation thereon. }
The views and conclusions contained herein are those of the authors and do not reflect the official policies or endorsements, either expressed or implied, of the Air Force Research Laboratory, the Naval Research Laboratory, the Department of Defense, or the U.S. Government.

%The views expressed are those of the authors and do not reflect the official policy or position of the Department of Defense or the US Government.

\ifCLASSOPTIONcaptionsoff
  \newpage
\fi


%% file: sections/reference.tex


%\nocite{*}
\bibliographystyle{IEEEtran}
\bibliography{jerrybibtex}

%% file: sections/biography.tex
\vfill